\documentclass{aastex61}

\begin{document}

\title{Gravitational Radiation from Binaries: A Pedagogical Introduction}

\author{Amir Jafari}
\email{elenceq@jhu.edu}
\affil{Johns Hopkins University, Baltimore, USA}

\begin{abstract}
This short note serves as an introduction to gravitational radiation through reviewing the inspiral-plunge transition phase in extreme mass ratio binaries. We study the relativistic motion of a compact object (CO) of mass $m$ around a massive black hole of mass $M\gg m$. The Kerr-Newman metric, effective potential for the general case of elliptical orbits, gravitational radiation, orbital energy and angular momentum of a coalescing CO in Kerr spacetime and gravitational wave frequency and signal to noise ratio are briefly reviewed. The main focus is on the transition from inspiral to plunge for a CO assuming that a test particle approach is plausible in the regime $m\ll M$ without appealing to a perturbative analysis. The effective potential is used to obtain the properties of the Innermost Stable Circular Orbit (ISCO) near which the adiabatic inspiral phase ends abruptly and the CO enters the plunge phase. For the transition phase, the effective potential is expanded in terms of parameters such as the radial (coordinate) distance from the ISCO and the deviation of particle's angular momentum from its value at the ISCO to obtain the equation of motion. The equations of motion, during the inspiral and transition phases, are joined numerically and the gravitational wave frequency, number of wave cycles and signal to noise ratio (SN) during the transition is obtained for circular/inclined as well as elliptical/inclined orbits. The limitations and inaccuracies of the current methods used to approach this problem is discussed. A short introduction to the fundamental concepts of General Relativity, in particular Einstein Field Equations is also provided in the Appendix.
\end{abstract}

\section{Introduction}

Coalescing black hole binary systems, in the Extreme Mass Ratio (EMR) regime, are one of the promising sources of gravitational radiation detectable by the Laser Interferometer Space Antenna (LISA). The EMR  is the most relevant regime for compact stars inspiraling toward massive black holes, e.g., in galactic nuclei. The small mass ratio, typically a CO of mass $m\sim M_{\odot}$ and a black hole of mass $M\gg m$, is translated into high frequencies within the LISA frequency band, $f\sim 10-10^4 \;Hz$, compared with the frequencies expected from the comparable mass binaries, which are expected to be detected by LIGO with the frequency band $f\sim 10^{-4}-1 \;Hz$. The non-zero mass of the inspiraling CO makes it different from a test particle, therefore demanding a perturbative approach. This has physical consequences. For example, unlike a test particle with a well-defined Innermost Stable Circular Orbit (ISCO), a CO does not have such a stable orbit. However, test particle approach can be used as an approximation, which should become better as the particle mass $m$ becomes much smaller than the mass of the black hole $M$. Likewise, the concept of ISCO, replaced by a transition regime for a CO, can be employed as an approximation in studying the properties of the transition from inspiral to plunge. The frequency band of LISA will be more sensitive to EMR regime, e.g., stellar- mass black holes, white dwarfs, and neutron stars falling into supermassive $m/M \sim 10^{-4}-10^{-8}$. For comparable mass binaries, potentially targeted by LIGO, other methods must be used which will not be discussed here.

The last several decades have seen many developments in the analytical as well as numerical studies in analyzing the inspiral and ring-down phases and the corresponding gravitational wavefronts. For circular orbits, the gravitational waves emitted by test particles coalescing toward a Kerr black hole can be computed using Teukolsky's (1973) perturbation formalism for Kerr metric. For circular orbits, the frequency of the gravitational wave $f$ has a simple relationship with the orbital angular velocity; $2\pi f = n\Omega$, with $n$ being a positive integer (Poisson 1993b). The dominant harmonic corresponds to $n = 2$. With a known relationship between the orbital energy $E$ and $\Omega$, one can calculate the rate of change of frequency $f$, that is $df/dt$, by calculating $dE/dt$, which is the rate that orbital energy is lost to to gravitational waves. For the adiabatic inspiral phase, the rate of energy loss, $dE/dt$ also called gravitational luminosity, can be calculated using the Post-Newtonian approximation, at least for systems with non-relativistic velocities $v\sim (M/r)^{1/2}$ where $M$ is the system mass; see \S 4.

The gravitational radiation from radial plunge of a test particle, moving on a geodesic, from infinity onto a Schwarzschild black hole has been studied extensively (see e.g., Davis et al. 1971; Davis et al. 1972; Nagar et al. 2007). The problem of a particle with initially non-zero angular momentum plunging to a Schwarzschild black hole has been considered by Detweiler and Szedenits and Oohara and Nakamura  using the perturbation formalism of Teukolsky (1973). An inspiraling CO, with non-zero angular momentum, toward a black hole under the radiation reaction will not however move on a geodesic. For a bound object, the elliptical orbit will be affected by the gravitational radiation, which tends to circularize the orbit (Ryan 1996). Consequently, the CO inspirals on a quasi-circular orbit. This phase ends abruptly with a transition to plunge near its Last Stable Orbit (LSO) (see e.g., Ori \& Thorne 2000; Buonanno \& Xu 2000, Nagar et al. 2007) and finally leads to the merger and ring-down. The typical amplitude for the gravitational wave radiated in the late stages of coalescing black holes is of order $ h \sim 10^{-22}$ out to $100 Mpc$. The corresponding frequencies in the $kHz$ range increase as the orbital period decreases as a result of the gravitational radiation (Lincoln \& Will 1990). 

The transition, from inspiral to plunge, has remained the most poorly understood phase of coalescence. An understanding of this phase can provide a sensitive probe of the innermost regions of black hole spacetimes since the orbit in this phase pass as near as possible to the hole itself (O'Shaughnessy 2003). Transition from inspiral to plunge, for equal-mass black hole binaries with quasi-circular orbits, has been studied numerically by computing gravitational wavefronts (e.g., see Sperhake et al. 2008). Buonanno \& Damour (2000) discussed the case of coalescing binary black holes with comparable masses. The considered quasi-circular configuration, implicitly assuming that the radiation back-reaction would circularized the otherwise elliptical orbits. It is in fact a common assumption that circular orbits remain circular under the adiabatic radiation reaction (Ryan 1996). The angular momentum loss, through gravitational wave emission, circularizes orbits faster than they shrink (Peters 1964; Apostolatos et al. 1993; Abbott et al. 2017). The fact that circular orbits remain circular under radiation reaction allows one to infer the evolution of the Carter constant $\cal{Q}$ from the changes in the orbital energy and angular momentum (Hughes 2001).

One of the earliest analytical studies of this phase, for circular orbits, was given by Ori \& Thorne (2000) in EMR regime. They discussed the duration and observability (through gravitational waves captured by LISA) of the transition from circular and equatorial inspiral to plunge of stellar-mass objects into supermassive spinning black holes. O'Shaughnessy (2003) extended this work to eccentric and equatorial orbits and calculated the transition time and estimated the probability for LISA to observe such a transitions. One implication is that there is no universal length for the transition, which depends almost randomly on initial conditions. Accordingly, Ori and Thorne?s results for quasi-circular orbits should be interpreted as an upper bound on the length of eccentric transitions involving similar bodies. For low-mass bodies ($m\sim 7 M_{\odot}$) the chance is less than 10\% (depending strongly on the astrophysical assumptions) for the LISA to detect a transition event with the signal to noise ratio of $S/N>5$. Sundararajan (2008) provided an approximate model for the trajectory of a compact object as it transitions from an adiabatic inspiral to a geodesic plunge in Kerr Spacetime. Instead of focusing on the determination of the transition time and probability of its detection by LISA, he focused on generating the particle's world line during the transition. This can be regarded as an extension of the approach taken by Ori \& Thorne (2000) to eccentric and inclined orbits. Similar to the latter work, Sundararajan (2008) approximated the equations of motion using a Taylor expansion of the geodesic equations about the LSO and subjecting them to evolving energy, angular momentum and the Carter constant. These equation are integrated numerically providing the radial and angular trajectories for a typical inclined/circular orbit and also for an inclined/eccentric orbit. These numerical results indicate that the transition time is correlated with the coefficient of the first term in the Taylor expansion of the radial potential, which is represented by $A_{1}$ in the present paper.

In \S 2, we briefly review the basic features of the motion, and gravitational radiation, of a CO in Kerr spacetime. In \S 3, the model of Ori \& Thorne (2003) for equatorial and circular orbits is discussed in some details based on the expansion of the effective potential near the ISCO. Following Sundararajan (2008), main lines of the similar approaches for circular/inclined orbits and eccentric/inclined orbits are discussed, respectively, in \S 4 and \S 5. We also apply the same methodology to charged black holes in \S 6. We discuss these results and their limitations in \S 7.

\section{ Kerr-Newman Spacetime}

The Kerr-Newman metric is given in the form of the following line element in Boyer-Lindquist coordinates $(t, r, \theta, \phi)$:

\begin{equation}\label{2}
ds^2=-\Big( {\Delta-a^2 \sin^2 \theta   \over  \rho^2 }  \Big)dt^2-2\omega\overline\omega^2dtd\phi+\overline\omega^2 d\phi^2+\rho^2 d\theta^2+{\rho^2\over \Delta} dr^2,
\end{equation}
where
\begin{equation}\nonumber
\Delta=r^2+a^2-2Mr+G^2=M^2(S^2+R^2+\overline G^2-2R),
\end{equation}
\begin{equation}\nonumber
\rho=(r^2+a^2\cos^2\theta)^{1/2}=M(R^2+S^2\cos^2\theta)^{1/2},
\end{equation}
\begin{equation}\nonumber
\Sigma=\Big[ (r^2+a^2)^2-a^2\Delta \sin^2\theta     \Big]^{1/2}
\end{equation}
\begin{equation}\nonumber
\overline\omega={\Sigma\over\rho}\sin\theta,
\end{equation}
\begin{equation}\nonumber
\omega={a(r^2+a^2-\Delta)\over \Sigma^2},
\end{equation}
\begin{equation}\nonumber
a={J\over M}=MS,
\end{equation}
\begin{equation}\nonumber
R={r\over M},
\end{equation}
\begin{equation}\nonumber
\overline g={g\over M},\;\overline Q={Q\over M},
\end{equation}
\begin{equation}\nonumber
G^2=g^2+Q^2,\;\overline G^2={G^2\over M^2}={g^2+Q^2\over M^2}.
\end{equation}

The outer horizon corresponds to $\Delta=0$ which gives $r_+=M+\sqrt{M^2-(a^2+g^2+Q^2)}$. The static limit corresponds to the root of $g_{tt}=0$, or $\Delta=a^2\sin^2\theta$, which has the solution $r_0=M+\sqrt{M^2-(a^2\cos^2\theta+g^2+Q^2)}$.

Consider a test particle, of mass $m$ and electric charge $q$, moving in the background spacetime of a Kerr-Newman black hole with mass $M$, spin $J=aM=M^2 S$, electric charge $Q$ and magnetic charge $g$ (we will be primarily concerned with the Kerr black holes, however, for the reference, the following expressions for the metric and Lagrangian are given for the general case with non-zero electric and magnetic charges). The Lagrangian reads

\begin{equation}\label{1}
{\cal{L}}={m\over 2} g_{\mu\nu} {dx^\mu\over d\tau}{dx^\nu\over d\tau}+qg_{\mu\nu}A^\nu{dx^\mu\over d\tau},
\end{equation}
where $A^\nu=(A_0, \bf{A})$ is the potential four vector given by

\begin{equation}\label{4potential}
A_\mu dx^\mu=-{1\over \rho^2}( Qr+g a \cos \theta) dt+{1\over \rho^2}[Q r a \sin^2 \theta+g(r^2+a^2)\cos\theta]d\phi.
\end{equation}

Using the Lagrangian, given by equation (\ref{1}), two constants of motion, energy $E$ and angular momentum $L$, can be obtained as
\begin{equation}\label{energy}
\pi_t={\partial{\cal{L}}\over\partial (dt/d\tau)   }=mg_{tt} {dt\over d\tau}+m g_{t\phi}{d\phi\over d\tau}+qA_0=p_t+q A_t=-E,
\end{equation}
and
\begin{equation}\label{Amomentum}
\pi_\phi={\partial{\cal{L}}\over\partial (d\phi/d\tau)   }=mg_{t\phi} {dt\over d\tau}+m g_{\phi\phi}{d\phi\over d\tau}+qA_3=p_\phi+q A_\phi=L.
\end{equation}

Here, $L=L_z$ is the angular momentum along the $z$ axis (the rotation axis of the black hole).

In what follows, for simplicity, we set the magnetic charge of the black hole to zero $g=0$. The third and fourth constants of motion are the particle's rest mass 
\begin{equation}\label{mass}
-m^2=g_{\mu\nu}p^\mu p^\nu,
\end{equation}
 and the Carter's constant;
\begin{equation}
{\cal{Q}}=p_\theta^2+\cos\theta^2[a^2(m^2-E^2)+L^2 /\sin^{2}\theta ].
\end{equation}
The Carter's constant can be combined by $L$ and $E$ in the form of the following constant 
\begin{equation}
{\cal{K}}={\cal{Q}}+(L-aE)^2.
\end{equation}

These constants of motion can be combined with the Kerr-Newman metric, given by equation (\ref{2}) with $g=0$, to get the equations of motions. If we write the particle's four-momentum as $p^\alpha=dx^\alpha/d\lambda$, the equations of motion read (Misner et al. 1973)

\begin{equation}\label{Theta}
\rho^2 {d\theta\over d\lambda}=\sqrt{\Theta},
\end{equation}
\begin{equation}\label{R}
\rho^2 {d r\over d\lambda}=\sqrt{{\cal{R}}},
\end{equation}

\begin{equation}\label{Phi}
\rho^2 {d\phi\over d\lambda}=-(aE-{L\over \sin^2\theta})+{a\over\Delta}{\cal{P}},
\end{equation}

\begin{equation}\label{T}
\rho^2 {dt\over d\lambda}=a(L-aE\sin^2\theta)+(r^2+a^2){{\cal{P}}\over\Delta},
\end{equation}

where $\Theta={\cal{Q}}-\cos^2\theta[a^2(m^2-E^2)+L^2/\sin^2\theta]$, ${\cal{R}}={\cal{P}}^2-\Delta[m^2 r^2+(L-aE)^2+{\cal{Q}}]$ and ${\cal{P}}=E(r^2+a^2)-La-e Q r$.

One comment is in order about the numerical integration of $\theta$ equation above. Note that ${\Theta}$ vanishes at the turning points $\theta_{min}$ and $\theta_{max}$ where $0\leq \theta_{min}\leq \theta_{max}\leq \pi$ (see e.g., Hughes 2000; Sundrarajan 2008). A parametrization of $\theta$ can resolve this difficulty by choosing 

\begin{equation}\label{DefChi}
z=\cos^2\theta=z_- \cos^2\chi,
\end{equation}

with

\begin{equation}
b(z-z_-)(z-z_+)=bz^2-z {{\cal{Q}}+L^2+a^2(m^2-E^2)\over m^2 }+{{\cal{Q}}\over m^2},
\end{equation}
and $b=a^2(m^2-E^2)/m^2$. The equation of motion, for $\theta$, becomes 

\begin{equation}\label{Chi}
{d\chi\over dt}={ \sqrt{b(z_+-z)} \over \gamma+a^2 E z(\chi)/m},
\end{equation}
where 
\begin{equation}
\gamma={E\over m}\Big[ {(r^2+a^2)^2\over \Delta}-a^2 \Big]-{2Mr a L\over m \Delta }.
\end{equation}

Note that in our notation $L=L_z$ is the $z$-component of the particle's angular momentum.

The time evolution of the azimuthal angle, $d\phi/dt$, is obtained simply by dividing equation (\ref{Phi}) by equation (\ref{T}):

\begin{equation}\label{AngVelEvo}
{d\phi\over dt}={ -(aE-{L\over \sin^2\theta})+{a\over\Delta}{\cal{P}} \over  a(L-aE\sin^2\theta)+(r^2+a^2){{\cal{P}}\over\Delta}}.
\end{equation}

\subsection{Effective Potential and Circular Orbits}

For radial motion, equations (\ref{Theta}) and (\ref{R}) can be solved for the energy $E$:

\begin{equation}
\alpha E^2-2\beta E+\gamma=0;\;\;\;E={\beta+\sqrt{\beta^2-\alpha\gamma}\over \alpha},
\end{equation}
where we have chosen the positive sign for the square root because otherwise, the four-momentum would point toward the past (see Misner et al. 1972). The coefficients are given by

\begin{equation}
\alpha=(r^2+a^2)^2-\Delta a^2\sin^2\theta>0\;(everywhere\;outside\;horizon),
\end{equation}
\begin{equation}
\beta=(La+qQr)(r^2+a^2)-La\Delta,
\end{equation}
\begin{equation}
\gamma=(La+qQr)^2-\Delta(L/\sin\theta)^2-m^2\rho^2\Delta-\rho^4\Big[(dr/d\lambda)^2+(d\theta/d\lambda)^2\Delta\Big].
\end{equation}

Turning points are found by setting $dr/d\lambda=0$. In the equatorial plane, $\theta=\pi/2$, and defining $\alpha_0\equiv \alpha(\theta=\pi/2)=(r^2+a^2)^2-\Delta a^2$ and $\gamma_0\equiv\gamma(\theta=\pi/2; dr/d\lambda=0)=(La+qQr)^2-L^2 \Delta -m^2r^2\Delta$, the minimum value of $E$, called the effective potential (see below), is given by

\begin{equation}
V={\beta+\sqrt{\beta^2-\alpha_0\gamma_0}\over \alpha_0}.
\end{equation}

The allowed regions for the particle (of energy $E$ at infinity) corresponds to $V(r)\leq E$. The turning points correspond to $dr/d\lambda=0$ or equivalently, $E=V(r)$. Stable circular orbits can be found by minimizing the effective potential, so they occur at points where $dV(r)/dr=0$ (see below).

In order to use a notation more common in the literature, we may simplify the expressions above and write the radial equation of motion and the effective potential explicitly in terms of the energy per particle mass $E/m$ and angular momentum per mass $L/m$. In particular, for a neutral particle around a Kerr black hole ($q=Q=g=0$), radial equation of motion reads
\begin{equation}\label{EofRM}
\Big({dr\over d\tau} \Big)^2+V_{eff}^2=\tilde E^2,
\end{equation}
where $\tilde E=E/m$ and the effective potential is given by
\begin{equation}\label{effectiveV}
V_{eff}^2=\tilde E^2-{1\over R^4}\Big[ [\tilde E(R^2+a^2/M^2)-\overline L a/M ]^2-(R^2-2R+a^2/M^2)[R^2+(\overline L-a\tilde E/M)^2]   \Big],
\end{equation}
with $\overline L=\tilde L/M=L/(Mm)$.

For circular orbits, as before, we set $dr/d\tau=0$ and $dV_{eff}/dr=0$. Solving the resultant equations simultaneously leads to the following expressions for the energy and angular momentum for circular orbits (Bardeen et al. 1972):

\begin{equation}\label{orbital-energy}
\tilde E\equiv {E\over m}={r^{3/2}-2M r^{1/2}\pm a M^{1/2}        \over   r^{3/4}( r^{3/2}-3M r^{1/2}\pm 2a M^{1/2}      )^{1/2} },
\end{equation}
and
\begin{equation}\label{orbital-momentum}
\tilde L\equiv {L\over m}={\pm M^{1/2}(r^2\mp 2aM^{1/2} r^{1/2}+a^2)      \over   r^{3/4}( r^{3/2}-3M r^{1/2}\pm 2a M^{1/2}      )^{1/2} },
\end{equation}
where the upper sign, which will be taken here, corresponds to the direct orbits (co-rotating with $L>0$), whereas the lower sign corresponds to retrograde orbits (counter-rotating with $L<0$). The denominator in the above expressions has a real value if $r^{3/2}-3M r^{1/2}\pm 2a M^{1/2}   \geq 0$. 

It is easy to check that setting $a=0$ leads to the familiar effective potential for the Schwarzschild solution $V_{eff}^2=1-2M/r+L^2 /(m^2r^2)-2ML^2/(m^2r^3)$. Thus the radial equation of motion then reads $(dr/d\tau)^2=E^2/m^2-1+2M/r-L^2/(m^2  r^2)+2ML^2/(m^2  r^3)$. 

Substituting orbital energy and angular momentum, given respectively by equations (\ref{orbital-energy}) and (\ref{orbital-momentum}) above, in equations (\ref{energy}) and (\ref{Amomentum}) gives the angular velocity $d\phi/dt$ for a circular orbit:
\begin{equation}
\Omega\equiv{d\phi/d\tau\over dt/d\tau}={\pm 1\over MR^{3/2}\pm a}={\pm M^{1/2} \over  r^{3/2}\pm a M^{1/2} },
\end{equation}
where the equality holds for photons, i.e., an orbit with infinite energy per unit mass. This photon orbit is the innermost limit for circular orbits of particles (Bardeen et al. 1972):

\begin{equation}
r_{ph}=2M\Big[1+\cos\Big({2\over 3}\cos^{-1}(\mp S) \Big)  \Big].
\end{equation}
Circular orbits outward of the photon radius $r>r_{ph}$, with $\tilde E\equiv E/m>1$, are unbound. Bound circular orbits exist for $r>r_{mb}$ where $r_{mb}$ is the radius of marginally bound circular orbit with $\tilde E=1$ (Bardeen et al. 1972):

\begin{equation}
r_{mb}=2M\mp a+2M^{1/2} (M\mp a)^{1/2}.
\end{equation}

Stability for a circular orbit is guaranteed by having $d^2V^2_{eff}/dr^2\leq 0$ or
\begin{equation}
r^2-6Mr\pm 8aM^{1/2} r^{1/2}-3a^2\geq 0.
\end{equation}

The turning points are the extrema of $V_{eff}$ and therefore are obtained by solving $dV_{eff}/dr=0$. The smallest radius for stable circular orbits with the stability condition $d^2V_{eff}/dr^2<0$ is given by

\begin{equation}
R_{isco}\equiv \Big({r\over M}\Big)_{isco}=3+Z_2\mp [(3-Z_1)(3+Z_1+2Z_2)]^{1/2},
\end{equation}
where $Z_1=1+(1-a^2/M^2)^{1/3}[(1+a/M)^{1/3}+(1-a/M)^{1/3}]$ and $Z_2=(3az^2/M^2+Z_1^2)^{1/2}$ (Bardeen et al. 1972; Ori \& Thorne 2000). It is easy to see that for $a=0$ (Schwarzschild solution), $R_{isco}=6$.

For the reference, we also write $V^2_{eff}(r)$, $\tilde E$ and $\tilde L$ for the circular/equatorial orbits in Reissner-Nordstr\"om spacetime:

\begin{equation}\label{orbital-energy-RN}
\tilde E_{RN}= {1-2M/r+Q^2/r^2    \over   (1-3M/r+2Q^2/r^2     )^{1/2} },
\end{equation}
and
\begin{equation}\label{orbital-momentum-RN}
\tilde L_{RN}=\pm {(Mr-Q^2)^{1/2}    \over   (1-3M/r+2Q^2/r^2 )^{1/2}     }.
\end{equation}
Effective potential and angular velocity read
\begin{equation}
V_{RN}^2=\Big(1-{2M\over r}+{Q^2\over r^2}\Big)\Big({\tilde L^2\over r^2}+1\Big),
\end{equation}
and
\begin{equation}
\Omega_{RN}\equiv{d\phi\over dt}=\pm \sqrt{{M\over r^3}-{Q^2\over r^4} }.
\end{equation}

Also, from equations (\ref{energy}) and (\ref{Amomentum}), one can write

\begin{equation}\label{dottRN}
\Big({dt\over d\tau}\Big)_{RN}={\tilde E\over 1-2M/r+Q^2/r^2},
\end{equation}

and
\begin{equation}\label{dotphiRN}
\Big({d\phi\over d\tau}\Big)_{RN}={\tilde L\over r^2\sin^2\theta}.
\end{equation}

And the ISCO (on the equatorial plane) is given by

\begin{equation}\label{ISCO-RN}
R_{isco}^{RN}=2+{\cal{N}}^{-1/3}\Big(4-3\overline Q^2+{\cal{N}}^{2/3} \Big),
\end{equation}
where 
\begin{equation}
{\cal{N}}=8+2\overline Q^4+\overline Q^2\Big(-9+\sqrt{ 5-9\overline Q^2 +4\overline Q^4} \Big).
\end{equation}

Note that $R=r_{isco}/M$ decreases as $\overline Q=Q/M$ increases and for $Q/M=1$, we find $R_{isco}=4$ (for a detailed study of neutral particle motion in Reissner-Nordstr\"om spacetime, see Pugiese et al. 2011). In passing, we also note that for the extremal Reissner-Nordstr\"om black hole, $Q=M$ or in SI units: $2r_Q\equiv 2Q\sqrt{G/(4\pi\epsilon_0 c^4)}=2GM/c^2\equiv r_S$.

\subsection{Radiation From Binaries}

Linearization of Einstein field equations (for an introduction see Appendix A), 

$$R^{\mu\nu}-g^{\mu\nu} R/2=8\pi T^{\mu\nu}$$

using a post-Minkowskian approximation 

$$g_{\mu\nu}=\eta_{\mu\nu}+h_{\mu\nu}$$

 leads to the wave equation $\Box h_{\mu\nu}=8\pi S_{\mu\nu} $ with energy momentum tensor $S_{\mu\nu}$ (which is given in terms of $T_{\mu\nu}$). Choosing a gauge condition $\partial_\nu h_\mu^\nu(x)=(1/2)\partial_\mu h^\nu_\nu(x)$, one finds plane gravitational waves $h_{\mu\nu}=e_{\mu\nu}\exp(ik_\lambda x^\lambda)+e^*_{\mu\nu}\exp(-ik_\lambda x^\lambda)$ with two independent polarizations; $e_{\mu\nu}$.\footnote{The gauge choice $\partial_\nu h_\mu^\nu(x)=(1/2)\partial_\mu h^\nu_\nu(x)$ doesn't exhaust the gauge freedom completely, since it only reduces independent components of symmetric tensor $e_{\mu\nu}$ from $10$ to $6$. In Transverse Traceless Gauge (TT), one also applies the conditions $h_{0i}=0$ for $i=1, 2, 3$ and $Tr (h)\equiv h_\mu^\mu=0$ to use up the full gauge freedom. These lead to $e_{0i}=0$ and $e_\mu^\mu=0$, which together with $\partial_\nu h_\mu^\nu(x)=(1/2)\partial_\mu h^\nu_\nu(x)$, for $\mu=0$, require $e_{00}=0$. Therefore, there are only two independent polarizations.} Far from a non-relativistic gravitational radiation source (e.g., a slowly rotating binary), we find 

\begin{equation}
\overline h_{ij}(t,{\bf{x}})={2\over r}{d^2Q_{ij}\over dt^2},
\end{equation}

where $\overline h_{\mu\nu}(t,{\bf{x}})= h_{\mu\nu}(t,{\bf{x}})-(1/2)\eta_{\mu\nu}h^\lambda_\lambda$ is the trace-reversed amplitude, which is the solution of $\Box \overline h_{\mu\nu}=-16\pi T_{\mu\nu}$, and the quadrupole moment $Q_{ij}$ is defined as

\begin{equation}
Q_{ij}=\int \rho(t,{\bf{x}}) x_i x_jdV.
\end{equation}
Here, integration is taken over the volume of the source, $\rho=T^{00}$ is the mass density of the source and the origin of the coordinate system is at the center of mass. For example, a contact binary system with two stars of mass $M$ and radius $R$ and period of motion $T$, one can easily calculate the quadrupole moment; $Q_{ij}=2MR^2$. Therefore $d^2Q_{ij}/dt^2=2MR^2/T^2$, and one finds $\overline h_{ij}=4MR^2/(rT^2)=MR^2\Omega^2/(\pi^2 r)$ where $\Omega$ is the angular velocity.

The radiation power or gravitational luminosity, $L=dE/dt$, is given by the famous quadrupole formula 
\begin{equation}
\Big({dE\over dt}\Big)_{GW}={1\over 5} \Big\langle {d^3\bar Q_{ij}\over dt^3}{ d^3\bar Q^{ij}\over dt^3} \Big\rangle,
\end{equation}
 where $\bar Q_{ij}=Q_{ij}-\delta_{ij} Q^k_k/3$. As an example, for circular orbits $R_1=R_2=R$ in an equal mass binary $M_1=M_2=M$ moving with angular velocity $\Omega$, $Q^{ij}=2Mx^i(t)x^j(t)$ with $x=R\cos \Omega t$ and $y=R\sin \Omega t$. One finds $dE/dt=(128/5)\Omega^6M^2R^4\propto v^{10}$, where $v=R\Omega$. This result, using the Kepler's third law, can also be written as 
 
 $$ dE/dt=(128/5)4^{1/3}(\pi M/T)^{10/3}.$$
 
 In a simple Newtonian picture of a circular binary with two stars of mass $M$ and separation $2R$, the total energy is given by $E_{tot}=Mv^2/2+Mv^2/2-2M/2R$, which using the Kepler's third law $R^3/T^2=M/16\pi^2$, yields $E_{tot}=-M^2/4R=-(M/4)(4\pi M/T)^{2/3}$. Since the system is bound, $E_{tot}<0$ and so radiation will decrease the period $T$. Equating the energy loss, $dE_{tot}/dt$, with the gravitational luminosity $dE/dt=(128/5)4^{1/3}(\pi M/T)^{10/3} $ gives us the change in period; 
 
 $$dT/dt=(-96\pi/5)4^{1/3}(2\pi M/T)^{5/3},$$
 
  which has been validated using pulsars. The loss of energy (and angular momentum) will shrink the orbit and also tend to circularize it if initially elliptical. The result is a quasi-circular configuration.

In Post-Newtonian approximation, the gravitational luminosity, $dE/dt$, for two masses $m_1$ and $m_2$ with corresponding spins ${\bf{s}}_1$ and ${\bf{s}}_2$, reduced mass $\mu$ and total mass $M$, is given by 
\begin{equation}\label{Luminosity}
\Big({dE\over dt}\Big)_{circ}={-32\over 5}\Big({\mu\over M}\Big)^2 v^{10} \Big[1-\Big( {1247\over 336}+{35\over 12}{\mu\over M}  \Big)v^2 +\Big(4\pi-{11\over 4}\zeta-{5\over 4}\xi  \Big)v^3+O(v^4)   \Big],
\end{equation}
where $\zeta=({\bf{s}}_1+{\bf{s}}_2).\hat{\bf{L}}$ and $\xi=\hat{\bf{L}}.( {m_2\over m_1}{\bf{s}}_1+{m_1\over m_2}{\bf{s}}_2)$. Here, $\hat{\bf{L}}$ is the unit vector in the direction of orbital angular momentum (Poisson 1993b). The terms inside the brackets in the above expressions are the Post-Newtonian corrections (Wagoner \& Will 1976; Poisson 1993a; Kidder et al. 1993). The term $(32/5)(\mu/M)^2v^{10}$ is the Newtonian result for circular orbits discussed above, which is generalized to elliptical orbits, with eccentricity $e$ and semimajor axis $a$, as (Peters \& Mattews 1963)
\begin{eqnarray}\nonumber
\Big({dE\over dt}\Big)_{N}&=&{-32\over 5}\Big({\mu\over M}\Big)^2 v^{10} \Big[1+{73\over 24}e^2+{37\over 96}e^4   \Big]\\
&=&{-32\over 5}{m_1^2m_2^2(m_1+m_2)\over a^5(1-e^2)^{7/2}}\Big[1+{73\over 24}e^2+{37\over 96}e^4   \Big],\label{Poisson-1993}
\end{eqnarray}
where we have used the definitions $\mu=m_1m_2/(m_1+m_2)$ and $M=m_1+m_2$.

\section{Equatorial and Circular Orbits}

One method to study the transition regime, and the properties of the radiated waves during this phase, is to expand the effective potential in terms of few small parameters. The idea is to approximate the effective potential, so the equation of motion, for the transitioning particle by Taylor expanding it in terms of small deviations in particle's orbital distance and angular momentum that particle acquires after leaving the ISCO. In this section, we elaborate this approach for circular orbits following Ori \& Thorne (2000).

Suppose a particle with mass $m$ moves on a quasi-circular orbit around a Kerr black hole with mass $M\gg m$, with the change in its radius $\Delta r$ supposedly being much smaller than its orbital radius, $\Delta r\ll r$. The change in particle's energy and angular momentum, because of gravitational radiation, are related as

\begin{equation}\label{EL}
{d\tilde E\over d\tau}=\Omega {d\tilde L\over d\tau}.
\end{equation} 

For a circular orbit ($e=0$), equation (\ref{Poisson-1993}) indicates that the test particle moving in the Kerr spacetime will lose energy via radiating gravitational waves with the rate 
\begin{equation}\label{32}
\dot E_{GW}=-\dot E={32\over 5} \Big({m\over M}\Big)^2 (M\Omega)^{10/3} \dot {\cal{E}},
\end{equation}

where $ \dot {\cal{E}}=d{\cal{E}}/dt$ is the general relativistic correction to the Newtonian quadrupole-moment expression given by the terms inside the brackets in equation (\ref{Luminosity}). Therefore the orbit gradually shrinks at a rate 
\begin{equation}
{dr\over dt}={-\dot E_{GW}\over dE/dr}.
\end{equation}

\begin{figure} [h]
\centering
\includegraphics[scale=.7]{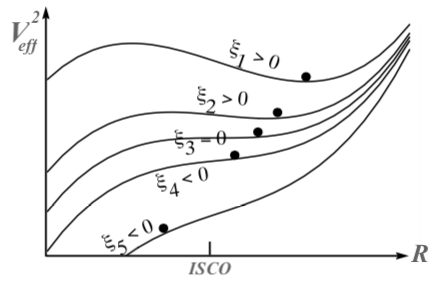}
\caption {\footnotesize {The effective potential for radial geodesic motion as a function of $\xi=\tilde L-\tilde L_{isco}$ (Ori \& Thorne 2000). Each curve corresponds to a specific value of $\xi$ which decreases as a result of radiation reaction. The particle, depicted as a large dot, initially is at the minimum of $V_{eff}$ ($\xi_1$; adiabatic regime) and reaches zero (near $\xi\simeq \xi_2$). The transition regimes ends at $\xi\simeq \xi_5$ and the particle plunges toward the black hole with almost constant energy and angular momentum (on a nearly geodesic trajectory).}}
\label{OriThorne2000}
\end{figure}

Near the ISCO, where the particle transitions to the plunge phase, equation (\ref{EL}) can be used to write

\begin{equation}\label{34}
\xi\equiv \overline L-\overline L_{isco}={\tilde E-\tilde E_{isco}\over M\Omega}.
\end{equation}
(Recall that in our notation, $\overline L=\tilde L/M=L/mM$.)
Assuming that $\tilde L_{isco}$ and $\tilde E_{isco}$ have known values, using the above expression, we can write the effective potential, equation (\ref{effectiveV}), in terms of these quantities 

\begin{eqnarray}\nonumber
V^2_{eff}(\xi)=(\tilde E_{isco}+M\Omega \xi)^2-{1\over R^4}&\Big[& [(\tilde E_{isco}+M\Omega \xi)(R^2+S^2)
-S(\overline L_{isco}+\xi)  ]^2\\
&-&(R^2-2R+S^2)\Big(R^2+\Big(\overline L_{isco}+\xi-S(\tilde E_{isco}+M\Omega \xi)\Big)^2\Big)  \Big].
\end{eqnarray}\label{effectiveV2}
This potential is plotted in Fig.(\ref{OriThorne2000}) for different values of $\xi$ (Ori \& Thorne 2000). For Reissner-Nordstr\"om black hole, we have
\begin{equation}
V^2_{eff}(\xi)=\Big(1-{2\over R}+{\overline Q^2\over R^2}\Big)\Big(1+{(\overline L_{isco}+\xi)^2\over R^2}\Big).
\end{equation}

In the extreme mass ratio regime, $m\ll M$, the gravitational radiation reaction is weak such that the particle, in the transition phase, has an angular velocity well approximated by the angular velocity of the ISCO, $\Omega_{isco}$ (Ori \& Thorne 2000):

\begin{equation}
\Omega\equiv {d\phi\over dt}\simeq\Omega_{isco},
\end{equation}

and 

\begin{equation}
 {d\tau \over dt}\simeq \Big( {d\tau \over dt}  \Big)_{isco}.
 \end{equation}

These expressions, for the ISCO, are given by (see e.g., Navikov \& Thorne 1973)

\begin{equation}
\Omega_{isco}={1\over M(S+R_{isco}^{3/2})},
\end{equation}

and

\begin{equation}\label{39}
 \Big( {d\tau \over dt}  \Big)_{isco}={ \Big(1-3/R_{isco}+2S/R_{isco}^{3/2} \Big)^{1/2}  \over 1+S/ R_{isco}^{3/2}}.
 \end{equation}
For Reissner-Nordstr\"om black hole, we have

\begin{equation}
\Omega_{isco}^{RN}={1\over M} \sqrt{{1\over R_{isco}^3}-{\overline Q^2\over R_{isco}^4} }.
\end{equation}

and

\begin{equation}\label{39RN}
\Big({d\tau\over dt}\Big)^{RN}_{isco}= (1-3/R_{isco}+2\overline Q^2/R_{isco}^2     )^{1/2}.
\end{equation}

The particle on quasi-circular motion near $r_{isco}$ radiates gravitational waves with energy and angular momentum approximately close to their values at $r_{isco}$. Combining equations (\ref{32}), (\ref{34}) and (\ref{39}), we find

\begin{equation}\nonumber
 {d\xi\over d\tau}={d\xi\over dt}{dt\over d\tau}=- {32\over 5}{m\over M^2}  (M\Omega_{isco})^{7/3} \dot{\cal{E}}_{isco}{  1+S/R_{isco}^{3/2}\over \Big(1-3/R_{isco}+2S/R_{isco}^{3/2} \Big)^{1/2} } .
\end{equation}\nonumber

If we choose $\tau$ such as $\xi(\tau=0)=0$, then 
\begin{equation}\label{xi}
\xi=- \Big[{32\over 5}  (M\Omega_{isco})^{7/3} \dot{\cal{E}}_{isco}{  1+S/R_{isco}^{3/2}\over \Big(1-3/R_{isco}+2S/R_{isco}^{3/2}\Big)^{1/2} }  \Big]{m\over M^2}\tau\equiv -\kappa {m\over M^2} \tau,
\end{equation}
where $\kappa$ is the term inside the brackets. 

For Reissner-Nordstr\"om black hole, we can similarly write

\begin{equation}\label{xi}
\xi^{RN}=- \Big[{32\over 5}  (M\Omega^{RN}_{isco})^{7/3} \dot{\cal{E}}^{RN}_{isco}(1-3/R_{isco}+2\overline Q^2/R_{isco}^2     )^{-1/2}  \Big]{m\over M^2}\tau\equiv -\kappa^{RN} {m\over M^2} \tau,
\end{equation}
where $\kappa^{RN}$ is the term inside the brackets in the first equation.

Once the particle is well into the transition regime, the radiation reaction becomes important. The equation of motion can be obtained by differentiating equation (\ref{EofRM});

\begin{equation}
{d^2r\over d\tau^2}=-{1\over2}{\partial V_{eff}\over \partial r}+F_{self},
\end{equation}
where $F_{self}$ is the self-force due to gravitational radiation reaction on the particle's motion, which will be ignored here (for a detailed discussion on this see Ori \& Thorne 2000). 

Assuming that we have full knowledge of the energy and angular momentum on the ISCO, we can expand the effective potential in terms of $\Delta R=R-R_{isco}$. To the first order in $\xi$, and second order in $\Delta R$, with a little tweak in notation, we re-write this expansion as
\begin{equation}\label{pot}
V^2_{eff}(\Delta R, \xi)={2A_1\over 3} \Delta R^3-2A_2\Delta R\xi+constant.,
\end{equation}
which leads to the equation of motion
\begin{eqnarray}\nonumber
{d^2\Delta R\over d\tau^2}&=&-A_1\Delta R^2+A_2 \xi\\\label{EofM10}
&=&-A_1\Delta R^2-A_2{m\over M^2}\kappa\tau,
\end{eqnarray}
where in the last line, we have used equation (\ref{xi}). The coefficients $A_1$ and $A_2$ are given by
\begin{eqnarray}\label{A1}
A_1&=&{1\over 4M^2}\Big({\partial^3 V^2_{eff}(R,\tilde E, \tilde L)\over \partial R^3} \Big)_{isco}\\
&=&{3\over M^2R_{isco}^6}\Big(R^2+2[S^2(\tilde E^2-1)-\overline L^2]R+10(\overline L-S\tilde E)^2\Big)_{isco},
\end{eqnarray}
and
\begin{eqnarray}
A_2&=&-{1\over 2M^2}\Big({\partial^2 V^2_{eff}(R, \tilde E, \tilde L)\over \partial \overline L\partial R}+M\Omega {\partial^2V^2_{eff}(R, \tilde E, \tilde L)\over \partial \tilde E\partial R}  \Big)_{isco}\\
&=&{2\over M^2R^4}\Big( (\overline L-S^2 \tilde EM\Omega)R-3(\overline L-S\tilde E)(1-SM\Omega)\Big)_{isco}.
\end{eqnarray}
which are equivalent to (3.18) and (3.19) in Ori \& Thorne (2000). 

For Reissner-Nordstr\"om black hole;

\begin{equation}
A_1^{RN}={3\over  M^2R^7}\Big( R(R^2+10\overline L^2-2R\overline L^2)-2\overline Q^2(R^2+5\overline L^2)\Big)_{isco},
\end{equation}
and
\begin{equation}
A_2^{RN}={-2\overline L\over M^2R^3}\Big[\Big({1\over R}-{\overline Q^2\over R^2}\Big)-\Big(1-{2\over R}+{\overline Q^2\over R^2}\Big) \Big]_{isco}.
\end{equation}

Using the parametrizations

\begin{equation}
X={R-R_{isco} \over(m/M)^{2/5}(A_2\kappa)^{2/5}A_1^{-3/5} },
\end{equation}
and

\begin{equation}\label{DefT}
T={\tau/M\over (m/M)^{-1/5} (A_1A_2\kappa)^{-1/5}},
\end{equation}

equation (\ref{EofM10}) can be cast into a dimensionless form 
\begin{equation}
{d^2X\over dT^2}=-X^2-T,
\end{equation}

which is equation (3.22) of Ori \& Thorne (2000). This is the (dimensionless) equation of motion for the transition regime, which should be smoothly connected to the equation of motion for the inspiral phase, equation (\ref{EofRM}), for $T\ll -1$. Since the ISCO is the circular orbit at the minimum of $V_{eff}$, therefore, using the potential (\ref{pot}), we get $\Delta R=(A_2\xi/A_1)^{1/2}$, which also can be written in the following form:

\begin{equation}
X=(-T)^{1/2},
\end{equation}

for adiabatic inspiral near the ISCO (Ori \& Thorne 2000). In the plunge phase, the particle moves, approximately, on a geodesic if we ignore the radiation reaction. This means the angular momentum as well as the energy of the particle remains almost constant, in other words $T\simeq 0$. Thus
\begin{equation}
{dX\over dT}=-\Big(constant-{2\over 3} X^3\Big)^{1/2}.
\end{equation}
For large $|X|$, one can neglect the constant term in the above equation and write

\begin{equation}
X=-6 (T_{plunge}-T)^{-2},
\end{equation}

for the plunge phase near the ISCO (Ori \& Thorne 2000). Using equations (\ref{34}) and (\ref{xi}), we can find the deficits in angular momentum and energy during the transition phase as 

\begin{equation}
\overline L_{final}-\overline L_{isco}=- \Big( \kappa (A_1A_2\kappa)^{-1/5} T_{plunge}\Big)\Big({m\over M}\Big)^{4/5},
\end{equation}
and
\begin{equation}
\tilde E_{final}-\tilde E_{isco}=- M\Omega_{isco} \Big( \kappa (A_1A_2\kappa)^{-1/5} T_{plunge}\Big)\Big({m\over M}\Big)^{4/5}.
\end{equation}

The transition solution is estimated by adiabatic inspiral at times $T < -1$. However, for $T > -1$ it deviates from adiabatic inspiral and evolves smoothly into a plunge. The solution diverges, namely $X \rightarrow - \infty$, at a finite time $T = T_{plunge}\simeq  3.412$ (Ori \& Thorne 2000).

\begin{figure} [h]
\centering
\includegraphics[scale=.75]{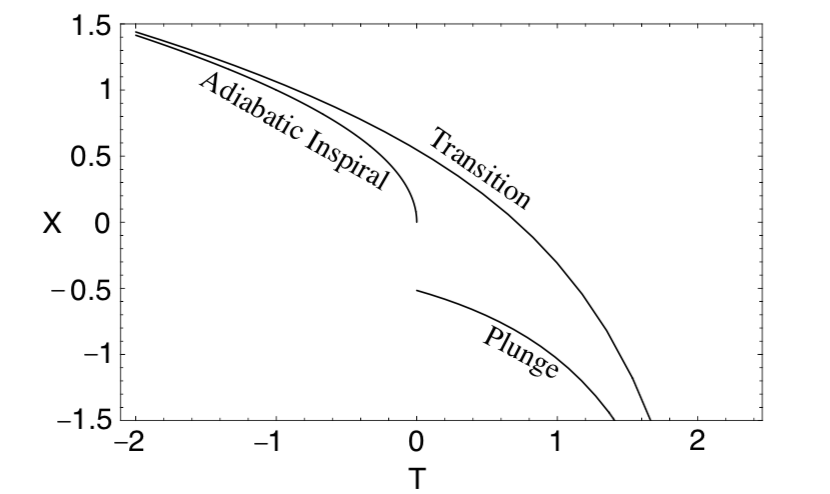}
\caption {\footnotesize {Dimensionless radius $X$ versus dimensionless proper time $T$ near the ISCO (Ori \& Thorne 2000). }}
\label{Thorne2}
\end{figure}

The equation of motion, solved numerically, is valid for $-1<T<2.3$ and $-5<X<1$, therefore $\Delta T=3.3$ and $\Delta X=6$. 

The frequency of the waves emitted during the transition regime would have a peak at

\begin{equation}
f \simeq 2 {\Omega_{isco}\over 2\pi}.
\end{equation}

The duration of the transition waves, detectable on Earth, is
\begin{equation}
\Delta t={M\over (d\tau/dt)_{isco}} \Big({m\over M} \Big)^{-1/5} (A_1A_2\kappa)^{-1/5} \Delta T,
\end{equation}
(This expression corresponds to equation (4.3) in Ori \& Thorne (2000) that has a missing $M$.) The frequency band $\Delta f=(1/\pi )(d\Omega /dR)_{isco}$;

\begin{equation}
\Delta f={3 M\over 2\pi} \Omega_{isco}^2 R_{isco}^{1/2}\Big({m\over M}\Big)^{2/5} (A_2\kappa)^{2/5}A_1^{-3/5} \Delta X.
\end{equation}

The number of gravitational wave cycles during the transition time is therefore given by

\begin{equation}
N_{cyc}=f \Delta t={\Omega_{isco} (A_1A_2\kappa)^{-1/5} \over \pi (d\tau/dt)_{isco)} } \Big( {m\over M}\Big)^{-1/5} \Delta T.
\end{equation}

The rms amplitude of the gravitational waves on Earth is

\begin{equation}
h^{rms}_{amp}=\Big\langle {h_+^2} _{amp} +{h_\times^2} _{amp} \Big\rangle^{1/2},
\end{equation}

where ${h_+} _{amp}$ and ${h_\times} _{amp}$ are the amplitudes for the second harmonic waves $h_+={h_+} _{amp} \cos (2\pi\int f dt+\phi_+)$ and $h_\times={h_\times} _{amp} \cos (2\pi\int f dt+\phi_\times)$. Therefore the power in the gravitational, detected at distance $D$ from the source, is given by

\begin{equation}
{dE\over dt}={4\pi \over 32\pi} D^2 (h^{rms}_{amp})^2(2\pi f)^2,
\end{equation}

which should be equal to the radiated power $(32\pi/5)(m/M)^2(M\Omega)^{10/3} \dot{\cal{E}}_{\infty,2}$. One finds

\begin{equation}
h^{rms}_{amp}={8\over \sqrt{5}}{m\over D}(M\Omega_{isco})^{2/3}\sqrt{\dot {\cal{E}}_{\infty,2} }.
\end{equation}

The signal to noise ratio (see Ori \& Thorne 2000; ) is

\begin{equation}
\Big({S\over N} \Big)_{rms}={h^{rms}_{amp}\over \sqrt{ 5S_h(f)/\Delta t}},
\end{equation}

where $5S_h(f)$ is the spectral density of LISA?s strain noise inverse-averaged over the sky and is given by (see )

\begin{equation}
S_h(f)=\Big[ (4.6\times 10^{-21})^2+(3.5\times 10^{-26})^2 \Big( {1\;Hz\over f}\Big)^4+(3.5\times 10^{-19})^2\Big({f\over 1\; Hz}\Big)^2\Big]\; Hz^{-1}.
\end{equation}

The numerical results are given in Table (\ref{OriThorneII}).

\begin{table} [h]
\centering
\includegraphics[scale=.6]{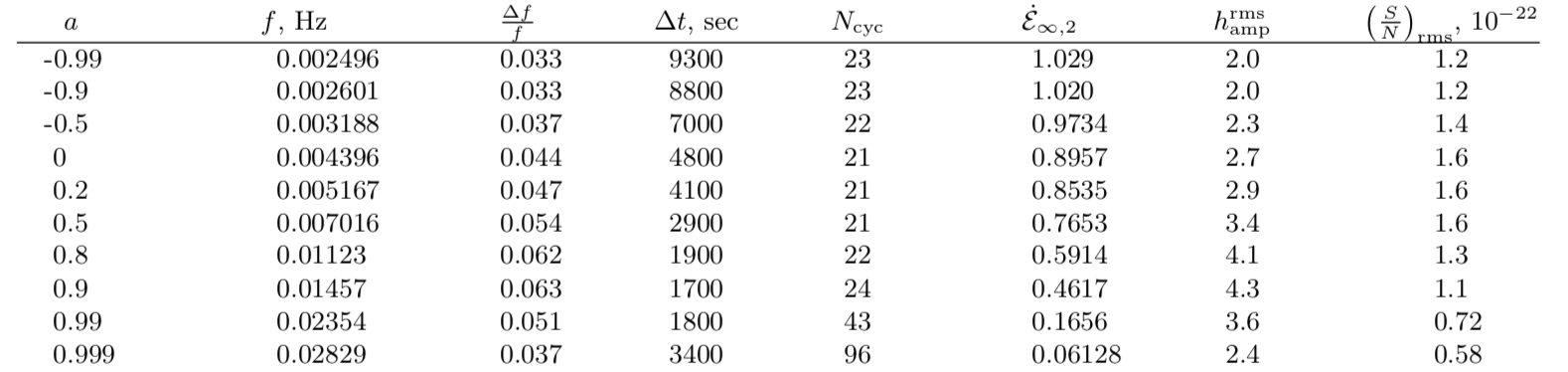}
\caption {\footnotesize {Properties of the second-harmonic,  transition-regime gravitational waves from a particle with mass $m=10 M_{\odot}$ inspiraling into a black hole of mass $M=10^6 M_{\odot}$ at distance $r=1\;Gpc$. From Ori \& Thorne (2000).}}
\label{OriThorneII}
\end{table}

\section{Inclined and Circular Orbits}

Inclination of an orbit can be defined by an angle $\iota$ in terms of the Carter constant $\cal{Q}$ and the component of the orbital angular momentum on the black hole's spin axis $z$;

\begin{equation}
\cos \iota={L\over \sqrt{L^2+{\cal{Q}} } }\equiv {L_z\over \sqrt{L_z^2+{\cal{Q}} } },
\end{equation}
which also represents our notation $L=L_z$.

To estimate the transitioning particle's angular momentum and energy, one may Taylor expand these quantities (Sundararajan 2008), around the LSO. Thus a generalization of equations (\ref{EL}) and (\ref{34}) for equatorial orbits, considered by Ori \& Thorne (2000), can be written as

\begin{equation}
E(t)\simeq E_{ISCO}+(t-t_{ISCO}) \dot E_{ISCO},
\end{equation}
\begin{equation}
L(t)\simeq L_{ ISCO}+(t-t_{ISCO}) \dot L_{ISCO},
\end{equation}
and
\begin{equation}
{\cal{Q}}(t)\simeq {\cal{Q}}_{ISCO}+(t-t_{ISCO}) (\dot {\cal{Q}}_{ISCO}+\dot {\delta {\cal{Q}}})+\delta {\cal{Q}},
\end{equation}

where, the supposedly small parameters $ {\delta {\cal{Q}}}$ and  $\dot {\delta {\cal{Q}}}$ are added to guarantee that the orbit remains circular near the transition. (For a detailed discussion of these equations and their underlying assumptions see Sundararajan (2008) and references therein.)

From equations of motion for $r$ and $t$, respectively given by (\ref{R}) and (\ref{T}), we can write

\begin{equation}
\Big( {dr\over dt}\Big)^2={{\cal{R}} \over \big(a(L-aE \sin^2 \theta)+(r^2+a^2){\cal{P}}/\Delta \big)^2   }\equiv {\cal{F}}(r, \chi),
\end{equation}

which, after taking derivative, leads to

\begin{equation}\label{Feq}
{d^2r\over dt^2}={1\over 2}\Big[ {\partial {\cal{F}}(r, \chi)\over \partial r}+{\partial {\cal{F}}(r, \chi)\over \partial \chi}{d\chi/dt\over dr/dt} \Big].
\end{equation}

where $\chi$ is defined by equation (\ref{DefChi}). Now, ${\cal{F}}(r,\chi)$ can be Taylor expanded (Sundararajan 2008) ignoring the terms of order $(m/M)^2$ and higher:

\begin{eqnarray}
{\cal{F}}(r, \chi, \iota, E, L)\simeq &{1\over 6}& {\partial^3 {\cal{F}}\over \partial r^3}|_{isco}(r-r_{isco})^3\\\nonumber
&+& {\partial^2{\cal{F}}\over \partial r\partial L}|_{isco} (r-r_{isco})(L-L_{isco})\\\nonumber
&+&  {\partial^2{\cal{F}}\over \partial r\partial E}|_{isco}(r-r_{isco})(E-E_{isco})\\\nonumber
&+&  {\partial^2{\cal{F}}\over \partial r\partial {\cal{Q}}}|_{isco}(r-r_{isco})({\cal{Q}}-{\cal{Q}}_{isco}).
\end{eqnarray}

Thus, equations (\ref{A1}) through (\ref{DefT}), for inclined orbits, become (Sundararajan 2008):

\begin{equation}\label{incA1}
A'_1=-{1\over 4}{\partial^3\over\partial R^3}\Big({ {\cal{R}}\over \rho^4}\Big)_{ISCO},
\end{equation}

\begin{equation}\label{incA2}
A'_2={1\over 2}\Big[ {\partial^2\over\partial \overline L\partial R}\Big({{\cal{R}}\over \rho^4}\Big)+{\dot {\tilde E} \over \dot {\overline L}} {\partial^2\over\partial \tilde E\partial R} \Big({{\cal{R}}\over \rho^4} \Big)+{\dot{\overline {\cal{Q}}}\over\dot {\overline L}}{\partial^2\over \partial \overline{\cal{Q}}\partial R} \Big({{\cal{R}}\over \rho^4} \Big)\Big]_{ISCO},
\end{equation}

\begin{figure} [h]
\centering
\includegraphics[scale=.69]{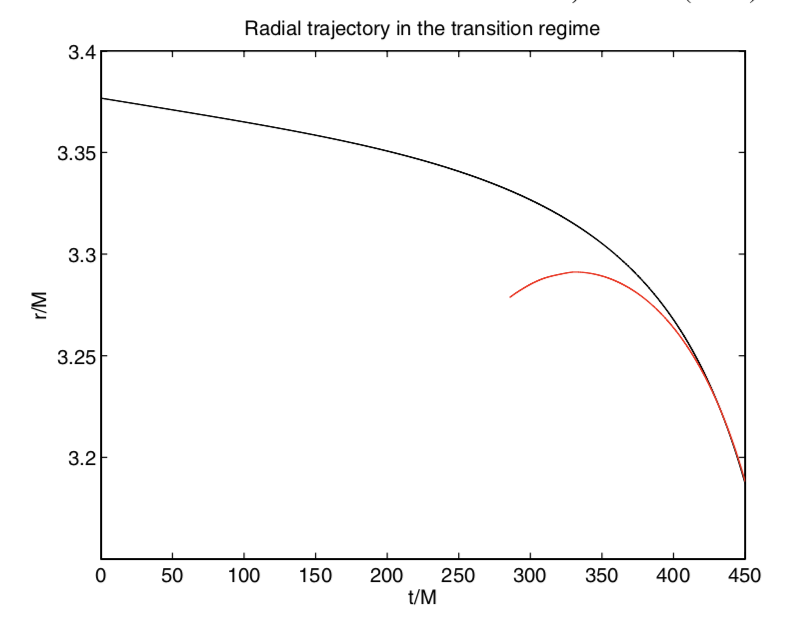}
\caption {\footnotesize {Radial trajectory during the transition (black line) from inspiral to plunge for a compact object of mass $m=10^{-5} M$ in a nearly circular orbit around a black hole with spin $a=0.8 M$ (from Sundararajan 2008). The compact object crosses the LSO at time $t_{lso}=137.5 M$. The inclination of the orbit at $t_{lso}$ is $\iota_{lso}=37^\circ$. The red (lower) line is a plunging geodesic matched to the end of the transition.}}
\label{Sundararajan2008}
\end{figure} 

\begin{equation}
X'={R-R_{isco} \over(m/M)^{2/5}(A_2\kappa_0)^{2/5}A_1^{-3/5} },
\end{equation}
\begin{equation}\label{DefT}
T'={t-t_{lso}\over (m/M)^{-1/5} (A_1A_2\kappa_0)^{-1/5}}{d\tau\over dt}|_{ISCO},
\end{equation}
where $\overline {\cal{Q}}={\cal{Q}}/(mM)^2$, $\kappa_0=\kappa_{lso}$ with $\kappa(t)$ defined as 

\begin{equation}
\kappa(t)=-{1\over m/M}{d\overline L\over d\tau/M}=-{Md\overline L/dt\over (m/M)(d\tau/dt)}.
\end{equation}
\begin{figure} [h]
\centering
\includegraphics[scale=.69]{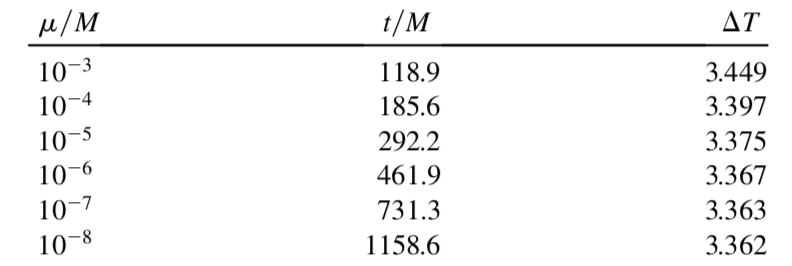}
\caption {\footnotesize {Variation of the transition time with $m/M$. Here $a=0.9 M$, $\iota_{lso}=0.001^\circ$, $M=1$, $T_s=-1$, $X_e=-5$ and also $r_{lso}=2.32 M$ (from Sundararajan 2008).}}
\label{Sundrarajan2008(2)}
\end{figure} 
Note that for inclined orbits $\iota\neq 0$, and unlike the previous section, $d\tau/dt$ varies with time since it is a function of $\theta(t)$. 

Sundararajan (2008) has evaluated $d\tau/dt$, $A_1'$, $A_2'$ and $\kappa_0$ at $\theta=\pi/2-\iota_{lso}$ (this choice comes from the fact that LSO in not known a priori). Following Ori \& Thorne (2008), as we did in the pervious section, one can set $T=-1$ at $t=0$ and numerically solve the equations up to $X\leq X_e=-5$. Sundararajan (2008) has also used the initial conditions $\phi=\chi=0$ (the latter being equivalent to $\theta=\theta_{min}$). The radial trajectory for the transition period is plotted in Fig. (\ref{Sundararajan2008}); for the numerical details see Sundararajan (2008).

\section{Inclined and Elliptical Orbits}

In this section, we follow the approach taken by Sundararajan (2008) for elliptical and inclined orbits. An elliptical orbit is represented by

\begin{equation}
r(t)={p\over 1+e \cos \psi},
\end{equation}

where $\psi(t)$, similar to the eccentric anomaly, is a function of time, $p$ is the semi-latus rectum, and $e$ is the eccentricity. The inner and outer turning points, $r_{min}$ and $r_{max}$, correspond, respectively, to $\psi=0, \pi$. The LSO is defined by

\begin{figure} [h]
\centering
\includegraphics[scale=.6]{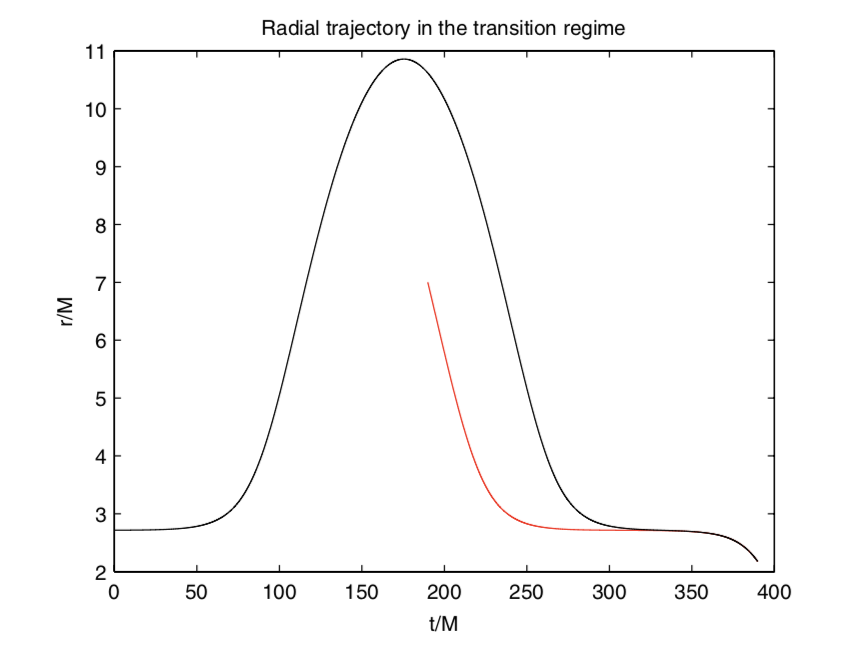}
\includegraphics[scale=.69]{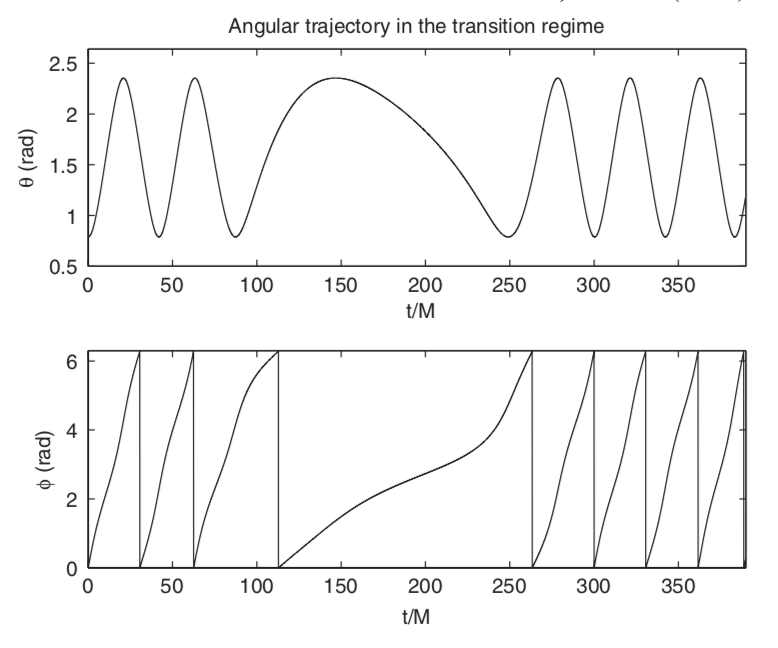}
\caption {\footnotesize {Top: Radial trajectory during the transition (black line) from inspiral to plunge for a compact object of mass $m=10^{-6}M$ in an eccentric orbit around a black hole with spin $a=0.8 M$. The compact object crosses the LSO at time $t_{LSO}=196.7M$. The inclination and eccentricity of the orbit at $t_{LSO}$ are $\iota_{LSO}=45^\circ$ and $e_{LSO}=0.6$. The red (lower) line is an unstable geodesic matched to the end of the transition.  Bottom: Angular trajectory during the transition for the same set of parameters as in Fig.(\ref{Sundararajan2008(3)}). Both plots from Sundararajan (2008).}}
\label{Sundararajan2008(3)}
\end{figure} 

\begin{equation}
{d{\cal{R}}\over dr}=0,\;\;\;at\;\;r=r_{min},
\end{equation}

and

\begin{equation}
{\cal{R}}=0,\;\;\;at\;\;r=r_{min} \;\&\;r=r_{max}.
\end{equation}

Similar to the inclined circular orbits, discussed in the previous section, we have

\begin{equation}
E(t)\simeq E_{LSO}+(t-t_{ISCO}) \dot E_{LSO},
\end{equation}
\begin{equation}
L_z(t)\simeq L_{z, LSO}+(t-t_{LSO}) \dot L_{z, LSO},
\end{equation}
and
\begin{equation}
{\cal{Q}}(t)\simeq {\cal{Q}}_{LSO}+(t-t_{LSO}) \dot {\cal{Q}}_{LSO}.
\end{equation}

The terms $\delta{\cal{Q}}$ and $\dot{\delta{\cal{Q}}}$ are omitted since there is no symmetry constraining ${\cal{Q}}(0)$ and $\dot{{\cal{Q}}}(0)$. So, in this case, $E(t)$, $L(t)$ and $\cal{Q}$ are independent (Sundararajan 2008). Similar to the circular case, the above expressions do not have the conservative effects of the self-force, and therefore, this leads to a slight shift of $(E, L< {\cal{Q}})_{LSO}$ and $p_{LSO}$ with respect to their geodesic values. 

To derive and solve the equations of motion for the general case of an inclined and elliptical orbit, one needs the time volution of the Carter constant; $\dot{\cal{Q}}$. 

One can still use equation (\ref{Chi}), $d\chi/ dt=\sqrt{b(z_+-z)} /( \gamma+a^2 E z(\chi)/m)$ and equation (\ref{AngVelEvo}) which has the following form:
\begin{equation}
{d\phi\over dt}={d\phi\over dt}(r, \chi).
\end{equation}

The geodesic equation is expanded at $(E_{LSO}, L_{LSO}, {\cal{Q}}_{LSO})$. Equation (\ref{Feq}) is used with

\begin{eqnarray}\nonumber
{d^2r\over dt^2}={1\over 2}\Big[ &&{1\over 2}{\partial^3{\cal{F}}\over \partial r^3}|_{LSO} (r-r_{LSO})^2+{\partial^2 {\cal{F}}\over \partial r\partial L}|_{LSO}(L-L_{LSO})^2\\\nonumber
&+&{\partial^2{\cal{F}}\over\partial r \partial E}|_{LSO}(E-E_{LSO})+{\partial^2{\cal{F}}\over \partial r\partial {\cal{Q}}}|_{LSO}({\cal{Q}}-{\cal{Q}}_{LSO})\\
&+&{\partial{\cal{F}}\over\partial \chi}{d\chi/dt\over dr/dt}\Big].
\end{eqnarray}

Fig.(\ref{Sundararajan2008(3)}) shows a typical trajectory during the inspiral-plunge transition for an eccentric orbit (see Sundararajan 2008 for details).

\begin{figure} [h]
\centering
\includegraphics[scale=.7]{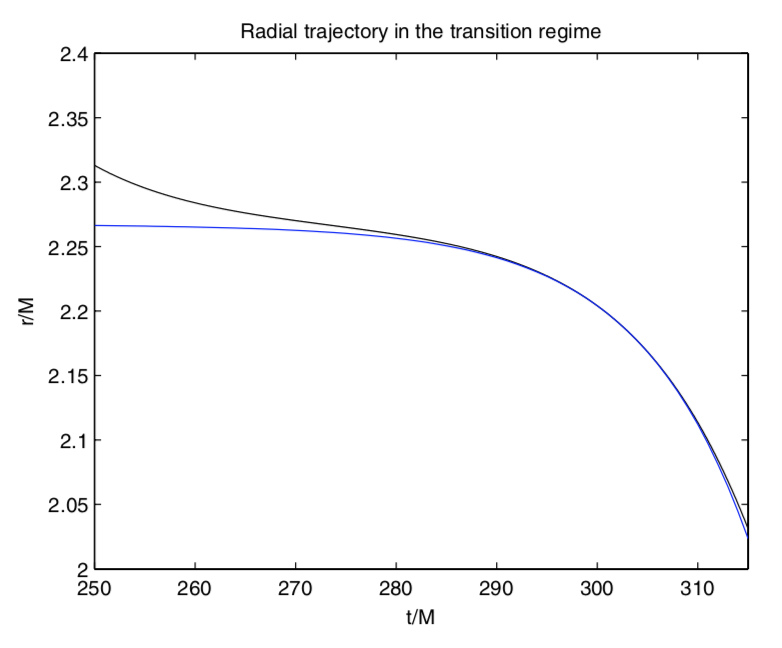}
\caption {\footnotesize {Comparison of the trajectories obtained by Sundararajan (2008), black curve, and O'Shaughnessy (2003), blue curve. The compact object is on an eccentric, equatorial orbit with parameters $e_{LSO}=0.6$ and $\mu=10^{-6} M$ around a black hole with unit spin $a=0.8 M$. From Sundararajan (2008).}}
\label{Sundararajan2008(5)}
\end{figure}

\section{Charged Black Holes}

Electric charge of black holes is usually assumed to be zero, or negligible. One way of testing this assumption is through calculating wave templates for gravitational waves from charged binaries. The inspiral-plunge transition phase of a Compact Object (CO) orbiting around an electrically charged, massive black hole, for example, generates gravitational waves intense enough to be potentially detectable by the Laser Interferometer Space Antenna (LISA). Here, we calculate the gravitational wave frequency, number of wave cycles, wave amplitude and signal to noise ratio for the transition phase of a CO moving on a circular and equatorial orbit around a massive charged black hole in the Extreme Mass Ratio regime. We compare these results with the Schwarzschild case, which might be used to measure the electric charge of the hole.

The metric $g_{\mu\nu}$ for an RN black hole of mass $M$ is given in the form of the following line element:

\begin{equation}\label{2}
ds^2=-{\Delta \over r^2} dt^2+ {r^2\over \Delta} dr^2+r^2 d\theta^2+r^2\sin^2\theta d\phi^2,
\end{equation}
where $\Delta=r^2-2Mr+Q^2$ and $Q$ is the electric charge of the black hole which produces the electric potential 

\begin{equation}\label{4potential}
A_0=-{Q \over r}.
\end{equation}
Adopting the notation $R=r/M$ and $\overline Q=Q/M$, the two horizons $r_{\pm}=M\pm(M^2-Q^2)^{1/2}$ can be cast into the more convenient dimensionless form $R_{\pm}=1\pm(1-\overline Q^2)^{1/2}$.

The motion of a neutral test particle of mass $m\ll M$, in RN spacetime, can be described using the Lagrangian
\begin{equation}\label{1}
{\cal{L}}={m\over 2} g_{\mu\nu} {dx^\mu\over d\tau}{dx^\nu\over d\tau}.\end{equation}

Two constants of motion, energy $E$ (measured by an observed at infinity) and angular momentum $L$, are
\begin{equation}\label{energy}
p_t={\partial{\cal{L}}\over\partial (dt/d\tau)   }=mg_{tt} {dt\over d\tau}=-E,
\end{equation}
and
\begin{equation}\label{Amomentum}
p_\phi={\partial{\cal{L}}\over\partial (d\phi/d\tau)   }=m g_{\phi\phi}{d\phi\over d\tau}=L.
\end{equation}

The third and fourth constants of motion are the particle's rest mass 

\begin{equation}\label{mass}
-m^2=g_{\mu\nu}p^\mu p^\nu,
\end{equation}
and the Carter's constant;
\begin{equation}
{\cal{Q}}=p_\theta^2+\cos\theta^2[L^2 /\sin^{2}\theta ].
\end{equation}
where $p^\alpha=dx^\alpha/d\lambda$ is the particle's four-momentum. The equations of motion read (Misner et al. 1973)

\begin{equation}\label{Theta}
r^2 {d\theta\over d\lambda}=p_\theta,
\end{equation}
\begin{equation}\label{R}
m r^2 {d r\over d\tau}=\sqrt{{\cal{R}}},
\end{equation}

\begin{equation}\label{Phi}
 m{d\phi\over d\tau}={L\over r^2\sin^2\theta},
\end{equation}

\begin{equation}\label{T}
 m{dt\over d\tau}={Er^2\over r^2+Q^2-2Mr}
\end{equation}

where $\Theta={\cal{Q}}-\cos^2\theta[L^2/\sin^2\theta]$, ${\cal{R}}=-(r^2+Q^2-2Mr)(m^2 r^2+L^2+{\cal{Q}})$.
We will restrict our study to equatorial ($\theta=\pi/2$) and circular orbits.

\subsection{Effective Potential and Circular Orbits}

\begin{figure} [h]
\centering
\includegraphics[scale=.55]{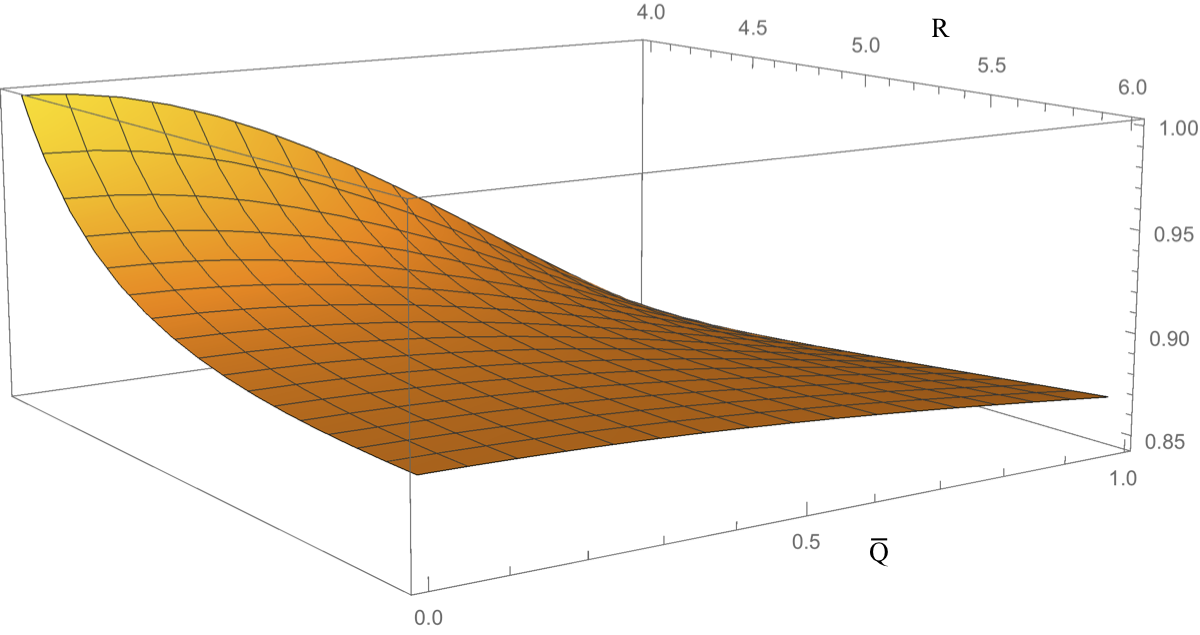}
\caption {\footnotesize {Effective potential $V^2_{eff}(R, \overline Q)$, given by eq.(\ref{Veff}), of a particle orbiting on the equatorial plane of a Reissner-Nordstr\"om black hole. }}
\label{Veff.png}
\end{figure}

The radial equation of motion, obtained combining (\ref{R}) and (\ref{T}), reads
\begin{equation}\label{EofRM}
\Big({dr\over d\tau} \Big)^2+V_{eff}^2=\tilde E^2,
\end{equation}
where $\tilde E=E/m$ and the effective potential $V^2_{eff}(r)$ is given by
\begin{equation}\label{Veff}
V_{eff}^2=\Big(1-{2\over R}+{\overline Q^2\over R^2}\Big)\Big({\overline L^2\over R^2}+1\Big),
\end{equation}
where $\overline L=\tilde L/M=L/(mM)$. This effective potential is plotted in Fig.(\ref{Veff.png}). 

For circular orbits, $dV^2_{eff}/dr=0$ and $\tilde E^2=V^2_{eff}$ (or equivalently $dr/d\tau=0$). Solving these equations simultaneously gives the particle's specific energy $\tilde E$ and specific angular momentum $\tilde L$:

\begin{equation}\label{orbital-energy-RN}
\tilde E=  {1-2/R+\overline Q^2/R^2    \over   (1-3/R+2\overline Q^2/R^2     )^{1/2} },
\end{equation}

\begin{equation}\label{orbital-momentum-RN}
\overline L=\pm {(R-\overline Q^2)^{1/2}    \over   (1-3/R+2\overline Q^2/R^2 )^{1/2}     }.
\end{equation}
Expressions (\ref{Phi}) and (\ref{T}) can be combined to obtain the angular velocity $\Omega=d\phi/dt=(d\phi/d\tau)(d\tau/dt)$, which upon substitution of $\tilde E$ and $\tilde L$, given above, yields
\begin{equation}
M\Omega\equiv M{d\phi\over dt}=\pm 1 \sqrt{{1\over R^3}-{\overline Q^2\over R^4} }.
\end{equation}

Eqs. (\ref{orbital-energy-RN}) and (\ref{orbital-momentum-RN}) require $R>\overline Q^2=R_*$ and $1-3/R+2\overline Q^2/R^2>0$. Thus motion is possible only for $R<R_{\gamma,-}$ and $R>R_{\gamma,+}$ where $R_{\gamma,\pm}=3\pm\sqrt{9-8\overline Q^2}$. Also, from eqs. (\ref{energy}) and (\ref{Amomentum}), one can write

\begin{eqnarray}\label{dottRN}
{dt\over d\tau}&=&{\tilde E\over 1-2M/r+Q^2/r^2}\\\nonumber
&=&(1-3/R+2\overline Q^2/R^2     )^{-1/2}.
\end{eqnarray}

The radius of the ISCO (where $V_{eff}^2=\tilde E^2$, $dV_{eff}^2/dR=d^2V_{eff}^2/dR^2=0$) is given by (Pugliese et al. 2011)

\begin{equation}\label{ISCO-RN}
R_{isco}=2+{\cal{N}}^{-1/3}\Big(4-3\overline Q^2+{\cal{N}}^{2/3} \Big),
\end{equation}
where 
\begin{equation}
{\cal{N}}=8+2\overline Q^4+\overline Q^2\Big(-9+\sqrt{ 5-9\overline Q^2 +4\overline Q^4} \Big).
\end{equation}

This radius is plotted in Fig.(\ref{R}). The gravitational radiation reaction will become important as particle leaves the ISCO and enters the plunge phase because of the energy lost to gravitational radiation.
\begin{figure} [h]
\centering
\includegraphics[scale=.4]{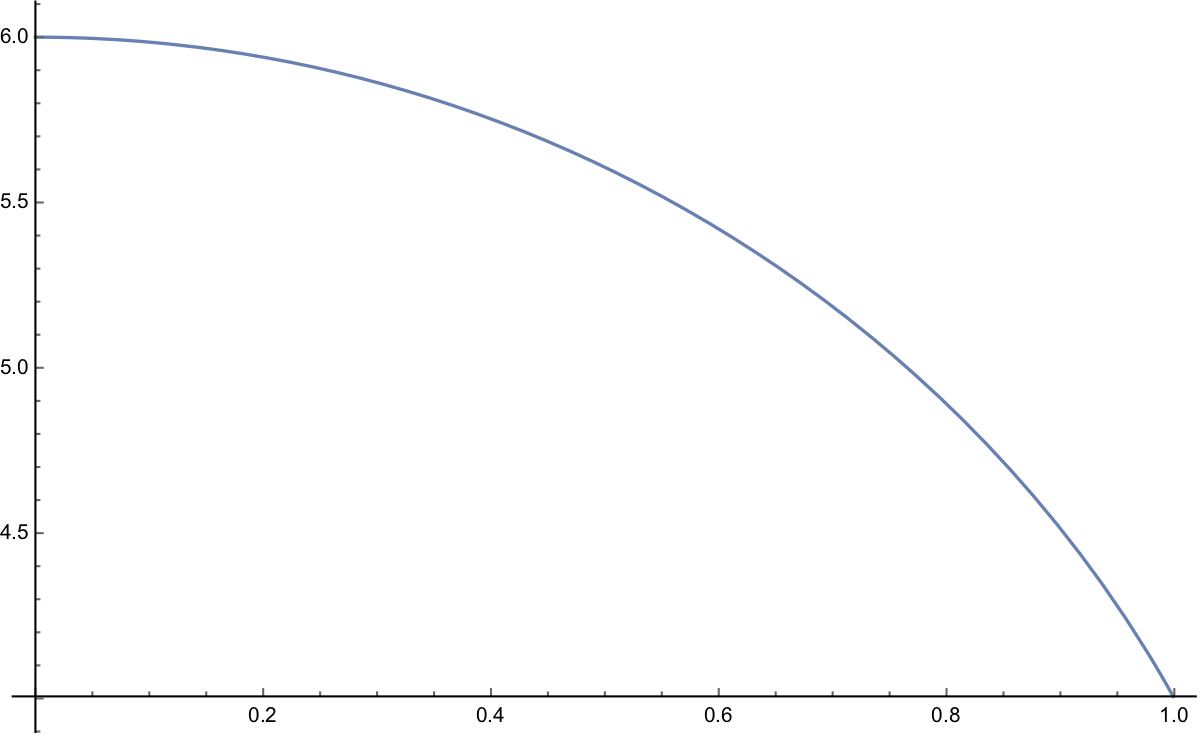}
\caption {\footnotesize {Dimensionless radius of the ISCO, $R_{isco}$, as a function of the specific charge of the black hole, $\overline Q=Q/M$. For $\overline Q=0$ (i.e., Schwarzschild case), we find $R_{isco}=6$ as expected. $R_{isco}$ decreases as $\overline Q$ increases and for $\overline Q=1$, we find $R_{isco}=4$.}}
\label{R}
\end{figure}

\subsection{Transition To Plunge}

In the EMR regime, the change in the particle's radius $\Delta R$ can be assumed to be much smaller than its orbital radius, $\Delta R\ll R$. Following Ori \& Thorne (2000), we write the change in the particle's energy and angular momentum, as a result of the gravitational radiation reaction, as 
\begin{equation}\label{EL}
{d\tilde E\over d\tau}=\Omega {d\tilde L\over d\tau}.
\end{equation} 

The particle on a circular orbit will lose energy via radiating gravitational waves with the rate 

\begin{equation}\label{32}
\dot E_{GW}=-\dot E\simeq {32\over 5} \Big({m\over M}\Big)^2 (M\Omega)^{10/3} ,
\end{equation}

which is the Newtonian quadrupole formula. The orbit gradually shrinks at a rate 
\begin{equation}
{dR\over dt}={-\dot E_{GW}\over dE/dR}.
\end{equation}

Near the ISCO, where the particle transitions to the plunge phase, eq.(\ref{EL}) can be used to write

\begin{equation}\label{34}
\xi\equiv \overline L-\overline L_{isco}={\tilde E-\tilde E_{isco}\over M\Omega}.
\end{equation}

The effective potential may now be written as

\begin{equation}
V^2_{eff}(\xi)=\Big(1-{2\over R}+{\overline Q^2\over R^2}\Big)\Big(1+{(\overline L_{isco}+\xi)^2\over R^2}\Big).
\end{equation}
Thus the particle's angular velocity can be approximated by its angular velocity at the ISCO, $\Omega_{isco}$, which is
\begin{equation}
M\Omega\simeq M\Omega_{isco}= \sqrt{{1\over R_{isco}^3}-{\overline Q^2\over R_{isco}^4} }.
\end{equation}

Also, from eq.(\ref{dottRN}), we have

\begin{equation}\label{39RN}
 {d\tau \over dt}\simeq \Big( {d\tau \over dt}  \Big)_{isco}= (1-3/R_{isco}+2\overline Q^2/R_{isco}^2     )^{1/2}.
\end{equation}

\begin{figure} [h]
\centering
\includegraphics[scale=.42]{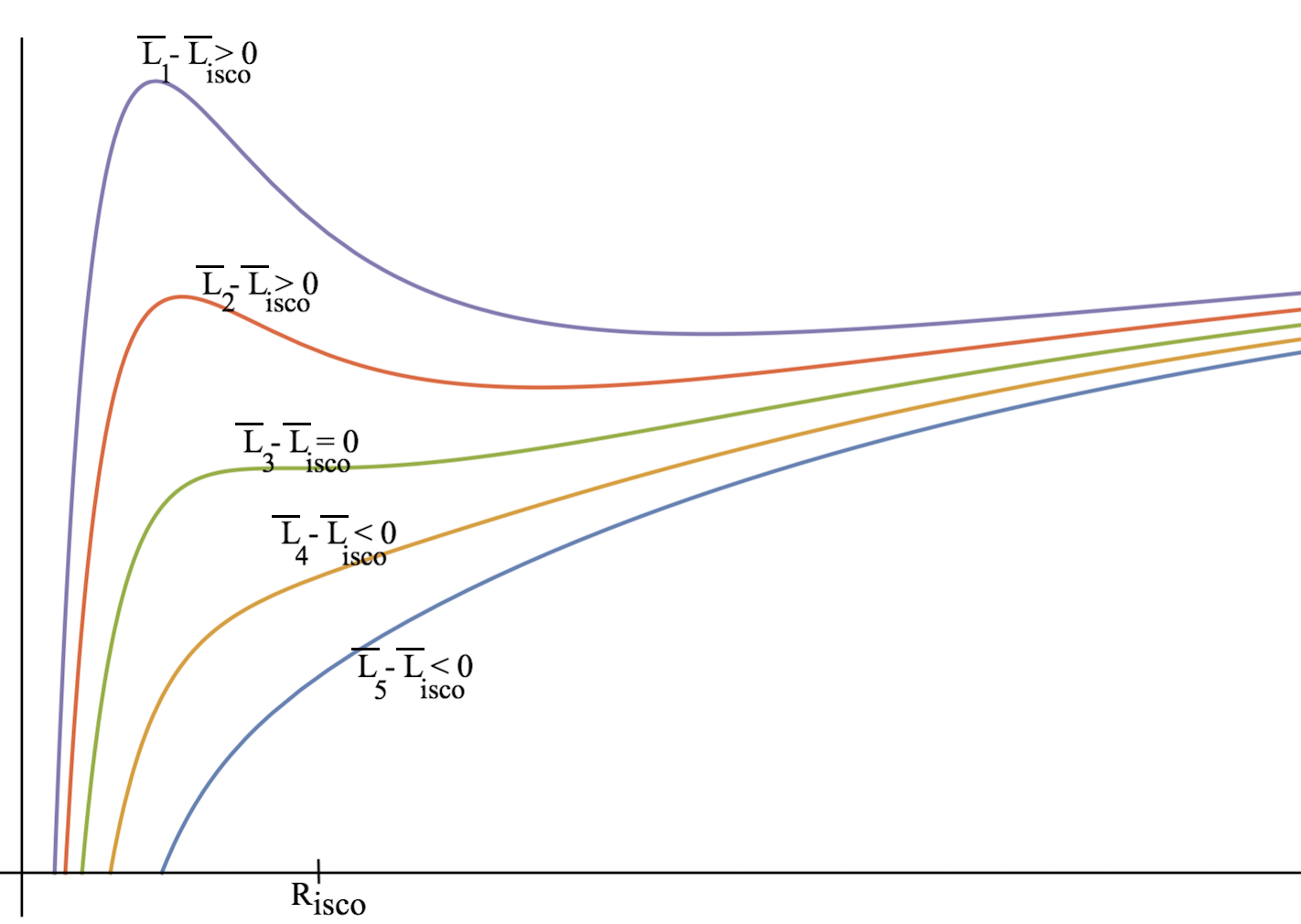}
\caption {\footnotesize {The effective potential plotted for different values of $\xi=\overline L-\overline L_{isco}$ with $\overline L_1>\overline L_2>\overline L_3>\overline L_4>\overline L_5$. For positive $\xi=\overline L-\overline L_{isco}$, the particle remains at the minimum of the potential. However, as $\xi$ decreases, due to the gravitational radiation reaction, the minimum of the potential goes inward and the particle cannot follow it and lags behind. At $\xi_4=\overline L_4-\overline L_{isco}$, the potential becomes so steep that transition ends and the particle plunges toward the black hole on a geodesic.  }}
\label{Potential}
\end{figure}

Consequently, the particle on quasi-circular motion near $R_{isco}$ radiates gravitational waves with energy and angular momentum approximately close to their values at $R_{isco}$, given respectively, by eq.(\ref{orbital-energy-RN}) and eq.(\ref{orbital-momentum-RN}). Combining eqs. (\ref{32}), (\ref{34}) and (\ref{39RN}), we find

\begin{equation}\nonumber
 {d\xi\over d\tau}={d\xi\over dt}{dt\over d\tau}=- {32\over 5}{m\over M^2}  (M\Omega_{isco})^{7/3} (1-3/R_{isco}+2\overline Q^2/R_{isco}^2     )^{-1/2}  .
\end{equation}\nonumber

If $\tau$ is chosen such as $\xi(\tau=0)=0$, then

\begin{equation}\label{xi}
\xi=- \Big[{32\over 5}  (M\Omega_{isco})^{7/3} (1-3/R_{isco}+2\overline Q^2/R_{isco}^2     )^{-1/2}  \Big]{m\over M^2}\tau\equiv -\kappa{m\over M} \tilde \tau,
\end{equation}
where $\tilde \tau=\tau/M$ and $\kappa=(32/5)  (M\Omega_{isco})^{7/3} (1-3/R_{isco}+2\overline Q^2/R_{isco}^2     )^{-1/2} $. 

The effective potential can be Taylor expanded (Ori \& Thorne 2000) in terms of $\Delta R=R-R_{isco}$ and $\xi$; the deviations of the particle's dimensionless (coordinate) radial distance and angular momentum from their values at the ISCO; see Fig.(\ref{Potential}). To the first order in $\xi$, and second order in $\Delta R$, this leads to

\begin{equation}\label{pot}
V^2_{eff}(\Delta R, \xi)={2A_1\over 3} \Delta R^3-2A_2\Delta R\xi+V_0^2,
\end{equation}
where $V_0$ is a constant. This expression gives the equation of radial motion in the transition regime
\begin{eqnarray}\nonumber
{d^2\Delta R\over d\tilde \tau^2}&=&-A_1\Delta R^2+A_2 \xi\\\label{EofM10}
&=&-A_1\Delta R^2-A_2{m\over M}\kappa\tilde \tau,
\end{eqnarray}
where in the last line, we have used eq.(\ref{xi}). The coefficients $A_1$ and $A_2$ are given by

\begin{equation}
A_1={3\over  R^7}\Big( R(R^2+10\overline L^2-2R\overline L^2)-2\overline Q^2(R^2+5\overline L^2)\Big)_{isco},
\end{equation}
and
\begin{equation}
A_2={-2\overline L\over R^3}\Big[\Big({1\over R}-{\overline Q^2\over R^2}\Big)-\Big(1-{2\over R}+{\overline Q^2\over R^2}\Big) \Big]_{isco}.
\end{equation}

Numerical values of these parameters are tabulated in Table.(\ref{table1}). Using the parametrizations

\begin{equation}
X={R-R_{isco} \over(m/M)^{2/5}(A_2\kappa)^{2/5}A_1^{-3/5} },
\end{equation}
and

\begin{equation}\label{DefT}
T={\tau/M\over (m/M)^{-1/5} (A_1A_2\kappa)^{-1/5}},
\end{equation}

eq.(\ref{EofM10}) is cast into a dimensionless form 
\begin{equation}\label{transition-phase}
{d^2X\over dT^2}=-X^2-T,
\end{equation}

\begin{table}[ht]
\caption{Dimensionless parameters characterizing the ISCO and the transition regime.}
\centering
\begin{tabular}{c c c c c c}
\hline\hline
$\overline Q$ & $R_{isco}$ & $M\Omega_{isco}$ & $A_1$ & $A_2$ & $\kappa$ \\ [0.5ex] 
\hline
0&6.000&0.06804&0.00077&0.01603&0.01710\\
0.1&5.984&0.06824&0.00077&0.1611&0.01723 \\
0.2&5.939&0.06884 &0.00079&0.01634&0.01763 \\
0.3 & 5.862 & 0.06990 & 0.00083&0.01675&0.01833 \\
0.4 & 5.752 & 0.07145 & 0.00089&0.01735&0.01941 \\ 
0.5 & 5.606 & 0.07362 & 0.00098&0.01821&0.02097 \\ 
0.6 & 5.419 & 0.07657 & 0.00110&0.01939&0.02322 \\ 
0.7 & 5.185 & 0.08619 & 0.00156&0.02338&0.03166 \\ 
0.8 & 4.890 & 0.08619 & 0.00156&0.02338&0.03166 \\ 
0.9 & 4.513 & 0.10643 & 0.00280&0.03231&0.05573 \\ 
0.99 & 4.060 & 0.10643 & 0.00280&0.03231&0.05573 \\ 
0.999 & 4.006 & 0.10806 & 0.00291&0.03306&0.05809 \\ [1ex]
\hline
\end{tabular}
\label{table1}
\end{table}

which is equation (3.22) of Ori \& Thorne (2000). This is the (dimensionless) equation of motion for the transition regime, which should be smoothly connected to the equation of motion for the inspiral phase, eq.(\ref{EofRM}), for $T\ll -1$. Since the ISCO is the circular orbit at the minimum of $V_{eff}$, therefore, using the potential (\ref{pot}), we get $\Delta R=(A_2\xi/A_1)^{1/2}$, which also can be written in the following form:

\begin{equation}\label{adiabatic-phase}
X=(-T)^{1/2},
\end{equation}

for adiabatic inspiral near the ISCO (Ori \& Thorne 2000). In the plunge phase, with radiation reaction neglected, the particle moves approximately on a geodesic. This means the angular momentum, as well as the energy, of the particle remains almost constant, in other words $T\simeq 0$. Thus
\begin{equation}
{dX\over dT}=-\Big(constant-{2\over 3} X^3\Big)^{1/2}.
\end{equation}
For large $|X|$, one can neglect the constant term in the above equation and write

\begin{equation}\label{plunge-phase}
X=-6 (T_{plunge}-T)^{-2},
\end{equation}

for the plunge phase near the ISCO (Ori \& Thorne 2000). Using eqs. (\ref{34}) and (\ref{xi}), we can find the deficits in angular momentum and energy during the transition phase as 

\begin{equation}
\overline L_{final}-\overline L_{isco}=- \Big( \kappa (A_1A_2\kappa)^{-1/5} T_{plunge}\Big)\Big({m\over M}\Big)^{4/5},
\end{equation}
and
\begin{equation}
\tilde E_{final}-\tilde E_{isco}=- M\Omega_{isco} \Big( \kappa (A_1A_2\kappa)^{-1/5} T_{plunge}\Big)\Big({m\over M}\Big)^{4/5}.
\end{equation}

The transition solution is estimated by adiabatic inspiral at times $T < -1$. However, for $T > -1$ it deviates from adiabatic inspiral and evolves smoothly into a plunge. The solution diverges, namely $X \rightarrow - \infty$, at a finite time $T = T_{plunge}\simeq  3.412$ (Ori \& Thorne 2000). The equation of motion, solved numerically, is valid for $-1<T<2.3$ and $-5<X<1$, therefore $\Delta T=3.3$ and $\Delta X=6$. Fig.(\ref{Thorne2}) shows the numerical solutions of equations of motion for the inspiral, transition and plunge phases plotted by Ori \& Thorne (2000).

\begin{figure} [h]
\centering
\includegraphics[scale=.7]{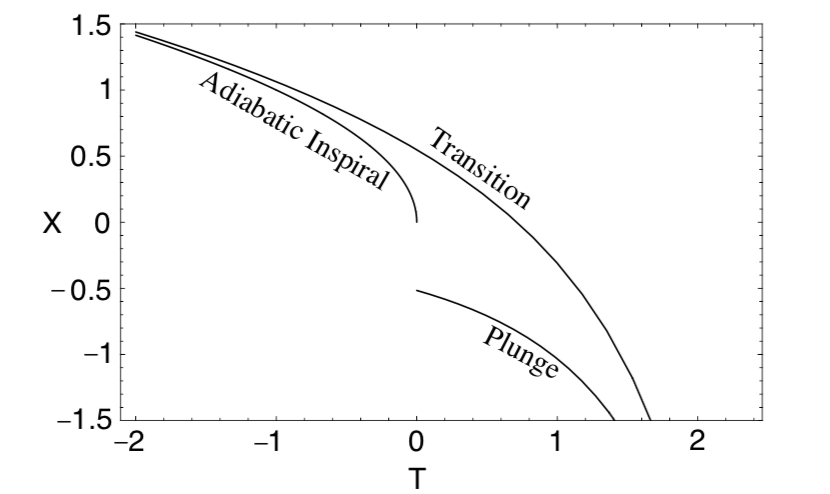}
\caption {\footnotesize {Dimensionless radius $X$ versus dimensionless proper time $T$ near the ISCO (Ori \& Thorne 2000). These solutions correspond to the equations of motion for the adiabatic inspiral phase, eq.(\ref{adiabatic-phase}), the transition phase, eq.(\ref{transition-phase}), and the plunge phase, eq.(\ref{plunge-phase}). }}
\label{Thorne2}
\end{figure}

\subsection{Gravitational Radiation}

The frequency of the waves emitted during the transition regime would have a peak at

\begin{equation}
f \simeq 2 {\Omega_{isco}\over 2\pi}.
\end{equation}

The duration of the transition waves, detectable on Earth, is
\begin{equation}
\Delta t={M\over (d\tau/dt)_{isco}} \Big({m\over M} \Big)^{-1/5} (A_1A_2\kappa)^{-1/5} \Delta T,
\end{equation}
and frequency band $\Delta f=(1/\pi )(d\Omega /dR)_{isco}$ is given by

\begin{equation}
\Delta f={3 M\over 2\pi} \Omega_{isco}^2 R_{isco}^{1/2}\Big({m\over M}\Big)^{2/5} (A_2\kappa)^{2/5}A_1^{-3/5} \Delta X.
\end{equation}

Therefore, the number of gravitational wave cycles during the transition time is 

\begin{equation}
N_{cyc}=f \Delta t={\Omega_{isco} (A_1A_2\kappa)^{-1/5} \over \pi (d\tau/dt)_{isco} } \Big( {m\over M}\Big)^{-1/5} \Delta T.
\end{equation}

\begin{table}[ht]
\caption{Frequency $f$, duration $\Delta t$, number of cycles $N=f\Delta t$, rms wave amplitude $h_{amp}^{rms}$ and signal to noise ratio $S/N$ from a compact object of mass $m=10 M_{\odot}$ around a charged black hole of mass $M=10^6 M_{\odot}$ at distance $D=10^9$ pc.}
\centering
\begin{tabular}{c c c c c c c}
\hline\hline
$\overline Q$ & $f $ (Hz)& $\Delta f/f $ & $\Delta t$ (s)& $N$ & $h_{amp}^{rms}$ ($\times 10^{-22}$) & $S/N$ \\ [0.3ex] 
\hline
0&0.004395&0.041&4971&22&2.9&1.7\\
0.01&0.004395&0.041&4971&22&2.9&1.7\\
0.1&0.004408&0.041&4953&22&2.9&1.7\\
0.2&0.004448&0.041&4900&22&2.9&1.7\\
0.3&0.004516&0.042&4812&22&2.9&1.7\\
0.4&0.004616&0.042&4686&22&3.0&1.8\\
0.5&0.004756&0.043&4523&22&3.0&1.8\\
0.6&0.004947&0.043&4318&21&3.1&1.8\\
0.7&0.005206&0.044&4070&21&3.2&1.8\\
0.8&0.005568&0.046&3772&21&3.3&1.8\\
0.9&0.006102&0.048&3413&21&3.6&1.8\\ 
0.999&0.006981&0.052&2979&21&3.9&1.8\\ [1ex]
\hline
\end{tabular}
\label{table2}
\end{table}
The rms amplitude of the gravitational waves on Earth is

\begin{equation}
h^{rms}_{amp}=\Big\langle {h_+^2} _{amp} +{h_\times^2} _{amp} \Big\rangle^{1/2},
\end{equation}

where ${h_+} _{amp}$ and ${h_\times} _{amp}$ are the amplitudes for the second harmonic waves $h_+={h_+} _{amp} \cos (2\pi\int f dt+\phi_+)$ and $h_\times={h_\times} _{amp} \cos (2\pi\int f dt+\phi_\times)$. Thus the power in the gravitational radiation, detected at distance $D$ from the source, is given by

\begin{equation}
{dE\over dt}={4\pi \over 32\pi} D^2 (h^{rms}_{amp})^2(2\pi f)^2,
\end{equation}

which should be approximately (in Newtonian approximation of the quadrupole formula) equal to the radiated power $(32\pi/5)(m/M)^2(M\Omega)^{10/3} $. One finds

\begin{equation}
h^{rms}_{amp}={8\over \sqrt{5}}{m\over D}(M\Omega_{isco})^{2/3}.
\end{equation}

The signal to noise ratio is

\begin{equation}
\Big({S\over N} \Big)_{rms}={h^{rms}_{amp}\over \sqrt{ 5S_h(f)/\Delta t}},
\end{equation}

where $5S_h(f)$ is the spectral density of LISA's strain noise inverse-averaged over the sky and is given by (Ori \& Thorne 2000)

\begin{equation}
S_h(f)=\Big[ (4.6\times 10^{-21})^2+(3.5\times 10^{-26})^2 \Big( {1\;Hz\over f}\Big)^4+(3.5\times 10^{-19})^2\Big({f\over 1\; Hz}\Big)^2\Big]\; Hz^{-1}.
\end{equation}

The numerical results are given in Table.(\ref{table2}), which summarizes the gravitational wave properties from a CO moving on the equatorial plane around a charged black hole and the properties of its radiated gravitational waves. A more general consideration, of course, would relax our restrictive assumptions of circular and equatorial motion and include a non-zero spin for the black hole. This is a straightforward but at the same time a cumbersome task given the complexity of such calculations for a Kerr-Newman black hole.

\section{Discussion}

Ori \& Thorne (2000) analytically approximated the transition between an adiabatic gravitational-wave inspiral and a plunge in extreme-mass-ratio inspirals in Kerr space-time. This computation provides a qualitatively correct picture but the quantitative details are not accurate because such calculations will receive corrections due to strong field dynamics of the radiation, the conservative piece of the self-force, and finite-size effects. These limitations were acknowledged or even expressed as unknown terms in the formulas of Ori \& Thorne (2000). The subsequent works, generalizing the results of Ori \& Thorne e.g., to non-circular orbits, essentially repeat the quantitatively inaccurate computation of Ori \& Thorne. Although these results may be insightful for future works but more accurate calculations are required to get a plausible picture. We have reviewed some of these generalizations in details in order to show their limitations and inaccuracies which may serve also as a pedagogical note. 

As for the charged black holes, it is well-known that astrophysical objects can acquire net electric charge through different processes. For instance, assuming thermal equilibrium, a larger number of electrons in the hot plasma on a stellar surface are expected to escape than much heavier ions. The generated strong electromagnetic fields, however, will not allow a large charge separation, keeping the acquired net charge on the star too small; about $100$ C for one solar mass (Bally \& Harrison 1977). For a black hole, as a collapsed star, of mass $M=10^6\;M_\odot$, this is translated into the ratio $\overline Q=(Q/M)(1/\sqrt{4\pi\epsilon_0 G})\sim 10^{-24}$. A comparison with the numerical results presented in Table.(\ref{table2}) shows that such a small charge-to-mass ratio would have an extremely negligible effect on the properties of the gravitational waves emitted by such systems. On the other hand, with recent developments in gravitational wave detection, it is desirable to test the assumption of zero (or negligible) electric charge on black holes in a direct way using observations. Nevertheless, the geometry of the Reissner-Nordström space-time is modified only with the square of the charge of the black hole, which, given the charges expected to accumulate in astrophysical black holes, makes any impact on neutral bodies absolutely negligible. Furthermore, the quantitative corrections named in the last paragraph will typically be many orders of magnitude larger than the contribution due to the black hole charge. As expected, a non-zero charge on the central hole affects the wave frequency and duration, radiated from an orbiting CO, only for charge-to-mass ratios considered to be unrealistically large in astrophysics. In any case, since appreciable charge-to-mass ratios are possible, at least in principle, it is of some interest to have wave templates and radiation parameters for such extreme cases even though the probability of observing such systems is small.

\appendix  \section{General Relativity}

Experiments have shown that the speed of light is indeed a constant. This requires the Lorentz transformations between two inertial frames:
\begin{equation}\label{GR-1}
dx'=\gamma \left( dx-{c\over c} cdt\right),\;
cdt'=\gamma \left(cdt-{v\over c}dx\right),
\end{equation}
where $\gamma=(1-v^2/c^2)^{-1/2}$. This form of relating the coordinates shows that spatial, $x$, and temporal, $t$, coordinates can be treated in the same way which is the essence of special relativity. To reflect this, and to emphasize on the equality of all inertial frames, we use four-vectors. We write
\begin{equation}
dx^{\mu}=(c dt, dx, dy, dz)\equiv (dx^0, dx^1, dx^2, dx^3),
\end{equation}
where $\;\mu=0, 1, 2, 3.$. The four velocity, four acceleration and four momentum can be written in a similar way:
\begin{equation}
U^\mu={dx^\mu\over d\tau}=\gamma (c, {\bf{v}}),\;
A^\mu={dU^\mu\over d\tau}=\left(\gamma^4    {  {\bf{a.}} {\bf{v}}\over c }, \gamma^2 {\bf{a}}+\gamma^4  {  {\bf{a.}} {\bf{v}}\over c^2 } {\bf{v}}\right),\;
P^\mu=mU^\mu,
\end{equation}
where $\tau$ is the proper time measured in the coordinate system moving with the particle; $d\tau=dt/ \gamma$, ${\bf{v}}={d{\bf{x}}\over dt}$ and ${\bf{a}}={d{\bf{v}}\over dt}$. The relativistic force is defined as $F^\mu=mA^\mu$ where $m$ is the rest mass of the particle.

The motion of an accelerated particle in the Minkowski space-time is described using Rindler coordinates:
\begin{equation}
ds^2=-\rho^2 d\sigma^2+d\rho^2,
\end{equation}
where $\rho=1/a$ and $\sigma=a \tau$ with $\tau$ being the proper time and $a$ the constant acceleration. Transformation tot he Cartesian coordinates is done using $x=\rho \cosh \sigma$ and $t=\rho \sinh \sigma$.

 Special relativity is based on the constancy of the speed of light and also the invariance of physical laws under uniform (constant speed) motion. It was natural to ask then about non-uniformly moving observers, that is observers with accelerated motion. But how was this related to gravity? Einstein realized the fact that falling objects are in the state of accelerated motion, they speed up as they fall. However, while in free fall, a falling person would not feel anything even his own weight. In other words, free fall (accelerated motion) somehow eliminates the effects of gravity. Einstein got this idea when he saw a worker falling from the roof through the window of his office in Bern: free fall (accelerated motion) eliminates gravity, Eureka!

To formulate gravity in a relativistic manner, we need to define physical laws in forms invariant under any arbitrary coordinate transformation. We know that tensors are quantities which preserve their properties under such transformations. So, a way to formulate a theory of gravity is to use tensors. To begin with, we define a four vector $A^\mu$ as a quantity that transforms as
\begin{equation}
A'^\mu={\partial x'^\mu \over \partial x^\nu}A^\nu,
\end{equation}
\begin{equation}
A'_\mu={\partial x^\nu \over \partial x'^\mu}A_\nu.
\end{equation}
The covariant derivative of the contravariant four vector $A^\mu$ is given by
\begin{equation}
DA^\mu=dA^\mu+\Gamma^\mu_{\alpha\beta} A^\alpha dx^\beta,
\end{equation}
and similarly for a covariant vector we write
\begin{equation}
DA_\mu=dA_\mu-\Gamma^\alpha_{\mu\beta} A_\alpha dx^\beta.
\end{equation}
The coordinate partial derivatives are usually written using a comma, and the covariant derivatives using a semicolon 
\begin{equation}
A^\mu_{\;,\nu}={\partial A^\mu\over \partial x^\nu}=\partial_\nu A^\mu,
\end{equation}
\begin{equation}
A^\mu_{\;;\nu}={DA^\mu\over \partial x^\nu}= A^\mu_{\;,\nu}+\Gamma^\mu_{\alpha\nu}A^\alpha.
\end{equation}
We can generelize this to the higher order derivatives writing $A^{\mu\alpha}_{\;;\beta}=D^2A^\mu/\partial x_\alpha\partial x^\beta$ (Peacock 1999). It is easy to see that the covariant derivative of a scalar is the ordinary gradient:
\begin{equation}
S_{;\mu}={\partial S\over \partial x^\mu}.
\end{equation}
The curl is also the same as the ordinary curl:
\begin{equation}
V_{\mu;\nu}-V_{\nu;\mu}={\partial V_\mu\over \partial x^\nu}-{\partial V_\nu\over \partial x^\mu}\equiv V_{\mu,\nu}-V_{\nu,\mu}.
\end{equation}
The reason for this general identity is the symmetry of the affine connection $\Gamma^\mu_{\alpha\beta}$ in $\alpha$ and $\beta$. The latter is also used to show that for an antisymmetric tensor $A^{\mu\nu}=-A^{\nu\mu}$:
\begin{equation}
A_{\mu\nu;\lambda}+A_{\lambda\mu;\nu}+A_{\nu\lambda;\mu}=A_{\mu\nu,\lambda}+A_{\lambda\mu,\nu}+A_{\nu\lambda,\mu}.
\end{equation}
The divergence is given by
\begin{equation}
V^\mu_{;\mu}={1\over \sqrt{g}}{\partial\over \partial x^\mu}\sqrt{g}V^\mu.
\end{equation}

We note that the covariant derivatives do not commute in general:
\begin{equation}
A^\mu_{\;;\alpha\beta}-A^\mu_{\;;\beta\alpha}=R^\mu_{\nu\beta\alpha}A^\nu,
\end{equation}
where $R^\mu_{\nu\beta\alpha}$ is called the Riemann-Christoffel curvature tensor:
\begin{equation}\label{Riemann-tensor}
R^\mu_{\nu\beta\alpha}={\partial\Gamma^\mu_{\nu\alpha}\over\partial x^\beta}-{\partial\Gamma^\mu_{\nu\beta}\over\partial x^\alpha}+\Gamma^\mu_{\sigma\beta} \Gamma^\sigma_{\alpha\nu}-\Gamma^\mu_{\sigma\alpha} \Gamma^\sigma_{\beta\nu}.
\end{equation}
The curvature tensor has the following properties (Dirac 1975):

$$R^\beta_{\nu\rho\sigma}=-R^\beta_{\nu\sigma\rho},$$
$$R^\beta_{\nu\rho\sigma}+R^\beta_{\rho\sigma\nu}+R^\beta_{\sigma\nu\rho}=0,$$
$$R_{\mu\nu\rho\sigma}=-R_{\nu\mu\rho\sigma},$$
$$R_{\mu\nu\rho\sigma}=R_{\rho\sigma\mu\nu}=R_{\sigma\rho\nu\mu},$$
$$R^\alpha_{\mu\rho\sigma;\tau}+R^\alpha_{\mu\sigma\tau;\rho}+R^\alpha_{\mu\tau\rho;\sigma}=0.$$

The last identity is known as the Bianchi identity which can be used to get a useful identity. By contraction we get

\begin{equation}
R_{\lambda\mu\nu\kappa;\eta}+R_{\lambda\mu\eta\nu;\kappa}+R_{\lambda\mu\kappa\eta;\nu}=0\rightarrow R_{\mu\kappa;\eta}-R_{\mu\eta;\kappa}+R^\nu_{\;\mu\kappa\eta;\nu}=0.
\end{equation}
A further contraction yields
\begin{equation}\label{einstein-tensor-1}
R_{\;\eta}-R^\mu_{\;\eta;\mu}-R^\nu_{\;\eta;\nu}=0\;\;or\;\;(R^{\mu\nu}-{1\over 2}g^{\mu\nu}R)_{;\mu}=0.
\end{equation}
The tensor $G^{\mu\nu}=R^{\mu\nu}-{1\over 2}g^{\mu\nu}R$ is called the Einstein tensor. The above equation then simply implies $G^{\mu\nu}_{;\mu}=0$.

Also, the Riemann tensor vanishes iff the space is flat. Contractions of this tensor give the Ricci tensor, and scalar or total curvature:
\begin{equation}\label{Ricci-tensor-1}
R_{\alpha\beta}=R^{\mu}_{\;\alpha\beta\mu}=R_{\beta\alpha},
\end{equation}
\begin{equation}\label{Ricci-tensor-2}
R=g^{\mu\nu}R_{\mu\nu}.
\end{equation}
The equation of motion of a particle in a gravitational field is just $DU^\mu/D\tau=0$ which we consider again below. In the presence of non-gravitational total four force $F^\mu$, the equation of motion becomes
\begin{equation}
m{DU^\mu\over D\tau}=F^\mu.
\end{equation}
The strong equivalence principle assumes that all the laws of physics have the same form in the freely falling frames as they would have in the absence of gravity (Peacock 1999). For example, the equation of motion in a freely falling system with special relativistic coordinates $\xi^\mu=(ct, x^1, x^2, x^3)$ and proper time $\tau$ is
\begin{equation}
{d^2 \xi^\mu\over d\tau^2}=0.
\end{equation}
This is the equation of a straight line in flat spacetime (Euclidean or in fact Minkowskian space). In this coordinate system, there is no acceleration and we can use the special relativity with the Minkowski meric;
\begin{equation}
c^2d\tau^2=\eta_{\alpha\beta}d\xi^\alpha d\xi^\beta,
\end{equation}
where $\eta_{\alpha\beta}=diag(1, -1, -1, -1)$. Going to other coordinate $x^\mu$ is done by using the transformation
\begin{equation}
d\xi^\mu={\partial d\xi^\mu \over \partial x^\nu}dx^\nu.
\end{equation}
Therefore, the equation of motion in the new coordinate system is
\begin{equation}
{d^2 x^\mu\over d\tau^2}+\Gamma^\mu_{\alpha\beta} {dx^\alpha\over d\tau}{dx^\beta\over d\tau}=0,
\end{equation}
with the new metric tensor $c^2d\tau^2=g_{\alpha\beta}dx^\alpha dx^\beta$. This is the equation of a geodesic; the shortest path in the curved spacetime (manifold). Here, we have
\begin{equation}
\Gamma_{\alpha\beta}^\mu={\partial x^\mu\over \partial \xi^\nu}{\partial^2\xi^\nu\over \partial x^\alpha \partial x^\beta},
\end{equation}
\begin{equation}
g_{\mu\nu}={\partial \xi^\alpha\over \partial x^\mu}{\partial \xi^\beta\over \partial x^\nu}\eta_{\alpha\beta}.
\end{equation}

The coefficients $\Gamma_{\alpha\beta}^\mu$ are Christoffel symbols, or affine connections, and are related roughly to the gravitational force whereas $g_{\mu\nu}$ is the new metric which can be thought as a symmetric matrix $\bf{g}$. So, there must be a coordinate system in which the metric is diagonal. The transformation that makes the matrix diagonal satisfies the following equation:
\begin{equation}
\tilde\Lambda {\bf{g}} \Lambda= diag(\lambda_0, . . . , \lambda_3),
\end{equation}
where $\Lambda$ is the matrix of transformation coefficients, and $\lambda_i$'s are the eigenvalues of this matrix. Differentiation of the metric yields
\begin{equation}
{\partial g_{\mu\nu}\over\partial x^\lambda}=\Gamma^\alpha_{\lambda\mu}g_{\alpha\nu}+\Gamma^\beta_{\lambda\nu}g_{\beta\mu},
\end{equation}
which can be used, by circulating the indices, to get an expression for the affine connections (Peacock 2010):
\begin{equation}
\Gamma^\alpha_{\lambda\mu}={1\over 2} g^{\alpha\nu}\left( {\partial g_{\mu\nu}\over \partial x^\lambda}+ {\partial g_{\lambda\nu}\over \partial x^\mu}- {\partial g_{\mu\lambda}\over \partial x^\nu}\right).
\end{equation}

If we use the relativistic notation, introduced so far, to re-write Maxwell equations, we should introduce the field tensor $F^{\mu\nu}=A_{\mu;\nu}-A_{\nu;\mu}$ where $A_\mu$ is the four potential vector.

We have the following relationships:

\begin{equation}
\begin{cases}
E_i=cF_{0i},\\
B_i=-{1\over 2}\epsilon_{ijk}F^{jk}.
\end{cases}
\end{equation}
and
\begin{equation}
\begin{cases}
{E^i\over c}=-F^{0i},\\
\epsilon^{ijk}B_k=-F^{ij}.
\end{cases}
\end{equation}

Maxwell's equations are written then
\begin{equation}
F^{\mu\nu}_{\;;\nu}=\mu_0J^\mu,
\end{equation}
\begin{equation}
{F_{\mu\nu}}_{;\gamma}+{F_{\nu\gamma}}_{;\mu}+{F_{\gamma\mu}}_{;\nu}=0.
\end{equation}
corresponds to the Faraday's and Gauss laws. Here, $J^\mu=(c\rho, {\bf{J}})$ is the four current which satisfies the continuity equation $J^\mu_{\;;\mu}=0$ with the electric charge density $\rho$. The latter is the relativistic form of $\partial_t \rho+{\nabla.}{\bf{J}}=0$. Now, if we seek a similar conservation law in mechanics, we should think of momentum and energy instead of the current. As we combined the charge density and current in one relativistic equation in electromagnetism, we do the same by combining momentum and energy in the energy-momentum tensor $\partial_\nu T^{\mu\nu}=0$. This analogy with electromagnetism (from Peacock 1999) helps to better understand the concept of the energy-momentum tensor: $T^{00}=c^2\times$(mass density)$=$energy density. $T^{12}=x$-component of the current of the $y$-momentum etc. In perfect fluid approximation, any point has a well-defined velocity, such that any observer moving with this velocity sees the fluid around as isotropic. This is important since we model the universe itself as a perfect fluid in cosmology. A requirement is for the mean free path between collisions to be small compared with the length scales in the fluid used by the observer (Weinberg 1972). For a frame in which the fluid is at rest. So, because of the fluid is assumed to be perfect, we get $T^{00}=c^2 \rho_0$ and $T^{ij}=P\delta_{ij}$ for $i,j=1, 2, 3$ where $P$ and $\rho$ are pressure and density. A Lorentz transformation to another arbitrary coordinate system gives the general form:
\begin{equation}\label{energy-momentum-perfect-fluid}
T^{\mu\nu}=(\rho+{P\over c^2})U^\mu U^\nu-Pg^{\mu\nu}.
\end{equation}
Note the sign convention in general relativistic equations: the above equation may be written using other sign conventions with a positive sign for the last term. The reason is the freedom in choosing any sign when defining (1) metric, (2) Riemann and (3) Einstein tensors.  We follow the notation adopted by Peacock 2010. The other convention mentioned above is, for example, adopted in Weinberg (1972).

The Einstein field equations govern the interaction of matter with the structure of the spacetime. They read
\begin{equation}\label{universe-1}
R_{\mu\nu}-{1\over 2}g_{\mu\nu}R=-{8\pi G \over c^4} T_{\mu\nu},
\end{equation}
where $R_{\mu\nu}$ and $R$ are the Ricci tensor and scalar respectively. Since the Einstein tensor $G_{\mu\nu}$ has $10$ independent components, we might think that the field equations give us $10$ independent differential equations which seem enough to find the metric with the same number of independent components. Unfortunately, the Bianchi identities relate the components of the Einstein tensor reducing the number of the algebraicly independent equations from $10$ to $6$. This means if $g_{\mu\nu}$ is a solution to the field equations, any other metric $g'_{\mu\nu}$ obtained from $g_{\mu\nu}$ using a coordinate transformation, which has $4$ arbitrary functions $x'^\mu(x)$, will be solution as well. This failure of the Einstein equations in determining metric is very similar to the gauge problem in classical electrodynamics: the Maxwell's equations as well cannot determine the vector potential uniquely. In the same way that we solve this problem in electrodynamics by fixing a gauge, we can adopt a particular coordinate system to get rid of the ambiguity in the metric  (Weinberg 1973). Here, similar to the Lorentz or Coulomb gauge fixing, we can use, for example, the harmonic coordinate conditions:
\begin{equation}
\Gamma^\lambda=g^{\mu\nu}\Gamma^\lambda_{\mu\nu}=0.
\end{equation}
The name harmonic coordinates comes from the fact that choosing $\Gamma^\lambda$ leads to the harmonic equation for the corresponding coordinates $\Box^2x^\mu =0$ with the d'Alembertian defined as
\begin{eqnarray}
\Box^2\phi&=&(g^{\lambda\kappa}\phi_{;\lambda})_{;\kappa}\\\label{dalembertian}
&=&g^{\lambda\kappa}{ \partial^2 \phi\over\partial x^\lambda\partial x^\kappa }-\Gamma^\lambda{\partial \phi\over \partial x^\lambda}.
\end{eqnarray}
If $\Gamma^\lambda=0$ then $\Box^2x^\mu =0$. If no gravitational field is present, the harmonic coordinate system is that of Minkowski (Weinberg 1973).

The energy-momentum tensor $T_{\mu\nu}$ in its general form is
\begin{eqnarray}
T^{\mu\nu}({\bf{x}})&=&\sum_i m {u_i}^{\mu}{u_i}^{\nu}\delta^3 ({\bf{x}}-{\bf{x}}_i)\\\label{universe-2}
&\equiv&\sum_i {c^2\over E_i} {p_i}^{\mu}{p_i}^{\nu}\delta^3 ({\bf{x}}-{\bf{x}}_i(t)),
\end{eqnarray}
where $E_i$ is the total energy (which is $\gamma m c^2$) and ${p_i}^{\mu}$ is the four-momentum of the $i$th particle. To get the conservation law, we take the derivative of the momentum-energy tensor as follows (Weinberg 1972):
\begin{eqnarray}
{\partial T^{\alpha i}\over \partial x^i}&=&\sum_n p_n^\alpha (t) {dx_n^i (t)\over dt} {\partial\over \partial x_n^i}\delta^3({\bf{x}}-{\bf{x}}_n(t))\\
&=&-\sum_n p_n^\alpha(t){\partial\over\partial t} \delta^3({\bf{x}}-{\bf{x}}_n(t))\\
&=&-{\partial\over \partial t} T^{\alpha 0}( {\bf{x}}, t)+\sum_n {dp_n^\alpha (t)\over dt} \delta^3({\bf{x}}-{\bf{x}}_n(t)).
\end{eqnarray}
We can re-write this result as a simple tensorial equation;
\begin{equation}
{\partial T^{\alpha i}\over \partial x^i}=G^\alpha,
\end{equation}
where $G^\alpha$ is the force density defined as
\begin{eqnarray}\label{universe-3}
G^\alpha({\bf{x}}, t)&=&\Sigma_n \delta^3({\bf{x}}-{\bf{x}}_n(t)) {dp_n^\alpha(t)\over dt}\\
&=&\sum_n \delta^3({\bf{x}}-{\bf{x}}_n(t)) f_n^\alpha {d\tau\over dt}.
\end{eqnarray}
For free particles $p_n^\alpha=const.$ and $G^\alpha=0$. The momentum energy tensor is not conserved, however, if there exists non-zero forces on the particles acting at a distance. Consider electromagnetic forces on charged particles $q_n$, for example. Then
\begin{eqnarray}
{\partial T^{\alpha \beta}\over \partial x^{\beta}}&=&\sum_n q_n F^\alpha_{\;\gamma}(x) {dx_n^\gamma\over dt} \delta^3({\bf{x}}-{\bf{x}}_n(t))\\\label{universe-4}
&=&F^\alpha_\gamma (x) J^\gamma(x),
\end{eqnarray}
where we have used the electromagnetic force written in terms of the electromagnetic tensor $F^{\mu\nu}$. Here $J^\mu$ is the foure current. We can resolve the problem by attributing a energy-momentum tensor to the electromagnetic field; $T_E^{\alpha\beta}$. Adding this to $T^{\alpha\beta}$ must give us $\partial_\beta (T^{\alpha\beta}+T_E^{\alpha\beta})=0$, so we must have $\partial_\beta T_E^{\alpha\beta}=-F^\alpha_\gamma (x) J^\gamma(x)$. The electromagnetic energy-momentum tensor is given by
\begin{equation}\label{universe-5}
T_E^{\alpha\beta}={1\over \mu_0} \left(F^\alpha_{\;\gamma}F^{\beta\gamma}-{1\over 4}\eta^{\alpha\beta}F_{\gamma\delta}F^{\gamma\delta}\right).
\end{equation}
Note that this tensor includes the Maxwell stress tensor $\sigma_{ij}$.

The classical Maxwell's tensor describes the interactions of matter and an electromagnetic field following the Lorents force density 
${\bf{f}}=\rho [ {\bf{E}}+{\bf{v}}\times{\bf{B}}]$ which can be re-written as
\begin{equation}\label{universe-7}
{\bf{f}}+\mu_0\epsilon_0{\partial {\bf{S}} \over \partial t}={\nabla.}{\sigma}.
\end{equation}
This is equivalent to $\partial_\beta (T^{\alpha\beta}+T_E^{\alpha\beta})=0$. In passing, we note that in addition to the above force equation, there is also an equation for energy:
\begin{equation}
{\partial u\over\partial t}+\nabla.{\bf{S}}+\bf{J.E}=0. 
\end{equation}
Both these force and energy equations are obtained from $T^{\alpha\beta}_{E,\beta}=G^\alpha$. Obviously, in a perfect fluid, like the universe, the total energy momentum tensor would be

\begin{equation}\label{energy-momentum-perfect-fluid2}
T^{\mu\nu}=(\rho+{P\over c^2})U^\mu U^\nu-Pg^{\mu\nu}+{1\over \mu_0} \left(F^\alpha_{\;\gamma}F^{\beta\gamma}-{1\over 4}g^{\alpha\beta}F_{\gamma\delta}F^{\gamma\delta}\right).
\end{equation}

Now, the total energy-momentum tensor is conserved. The need for adding an extra term to the energy-momentum tensor simply means that the electromagnetic field carries both energy and momentum which are taken care of using the electromagnetic energy-momentum tensor.

We can use $T^{\beta\gamma}$ to define another tensor (Weinberg 1972): $M^{\gamma\alpha\beta}=x^\alpha T^{\beta\gamma}-x^\beta T^{\alpha\gamma}$ which is conserved:
\begin{equation}
{\partial M^{\gamma\alpha\beta}\over \partial x^\gamma}=0.
\end{equation}

Next, we define the angular momentum tensor:
\begin{equation}
J^{\alpha\beta}=\int d^3x M^{0\alpha\beta}=-J^{\beta\alpha}.
\end{equation}
Since the angular momentum tensor includes the orbital momentum so it is not invariant under a translation $x^\alpha \rightarrow x^\alpha+a^\alpha$. The internal part of the angular momentum tensor is called spin:
\begin{equation}
S_\alpha={1\over2}\epsilon_{\alpha\beta\gamma\delta} J^{\beta\gamma} U^\delta;\;\;\;\;S_\alpha U^\alpha=0.
\end{equation}
For a free particel we get $dS_\alpha/dt=0$.

The field equations are generalization of the Poisson equation for the gravitational potential: $\nabla^2 \phi=4\pi G\rho$ (or in terms of metric $\nabla^2 g_{ll}=-8\pi GT_{ll}$). Note that in the LHS of the field equation, the Einstein tensor $G_{\mu\nu}=R_{\mu\nu}-{1\over 2}g_{\mu\nu}R$ describes the geometry of the spacetime while in the RHS, the stress-energy tensor $T_{\mu\nu}$ describes the distribution of the matter and energy.

It is important obviously that any theory must yield the known special cases which are already established. For the theory of general relativity, we must be able to re-derive the classical equations of motion used in Newtonian mechanics. Here, we derive the classical limit of the Einstein field equations and the equation of motion (Peacock 2010; Weinberg 1972). For events with the same spatial coordinates, the difference in proper time is
\begin{equation}\label{Newtonian-limit-1}
d\tau=dt\left(1+{\Delta\phi\over c^2}\right),
\end{equation}
where $\phi$ is the gravitational potential ${\nabla}\phi=-\bf{g}$. The line element is
\begin{equation}\label{Newtonian-limit-2}
d\tau^2=dt^2\left(1+{\Delta\phi\over c^2}\right)^2\equiv g_{00}dt^2.
\end{equation}
Therefore, we can write $g_{00}\simeq 1+2\Delta\phi/c^2$ since we are in the weak field approximation. This yields ${\nabla}g_{00}=(2/c^2){\nabla}\phi=-(2/c^2)\bf{g}$. We get
\begin{equation}\label{Newtonian-limit-3}
{\bf{g}}=-{c^2\over 2} {\nabla}g_{00}.
\end{equation}
The equation of motion becomes
\begin{equation}\label{Newtonian-limit-4}
{\ddot{x}}^i+c^2\Gamma^i_{00}=0,
\end{equation}
where $\Gamma^i_{00}=(g^{\nu i}/2)\left(0+0-\partial g_{00}/\partial x^\nu \right)$. This is the same as $\Gamma^i_{00}=(1/2){\nabla}g_{00}$. So, we get
\begin{equation}\label{Newtonian-limit-5}
{\ddot{x}}^i=-{c^2\over 2} {\nabla}g_{00}=\bf{g},
\end{equation}
which is familiar from classical mechanics.
To see how the field equations become the familiar Poisson's equation for the gravitational field, we look at their spatial part $R^{ij}-(1/2)g^{ij}R=-(8\pi G/C^4)T^{ij}$ which upon contraction gives $R=-2R^{00}-3P(16\pi G/c^4)$ using $g_{ij}R^{ij}=R-R^{00}$ and $g_{ij}T^{ij}=-3P$. This helps us to get the temporal part of the Einstein tensor as $G_{00}=G^{00}=2R_{00}+3P(8\pi G/c^4)$. To get $R_{00}$ we ignore the non-linear terms in the Riemann tensor to write 
\begin{equation}
R_{\alpha\beta}={\partial \Gamma^\mu_{\alpha\mu}\over \partial x^\beta}-{\partial \Gamma^\mu_{\alpha\beta}\over \partial x^\mu} \;\;\;\rightarrow R_{00}=-\Gamma^i_{00,i}={1\over 2}\nabla^2 g_{00}=-{1\over c^2}\nabla^2\phi.
\end{equation}
So, $G_00=-(8\pi G/c^4)T_{00}$ becomes
\begin{equation}
-{2\over c^2}\nabla^2\phi+3P\left ( {8\pi G\over c^4}\right)=-{8\pi G\over c^4}\rho c^2.
\end{equation}
This gives us the following equation for a perfect fluid 
\begin{equation}
\nabla^2\phi=4\pi G (\rho+{3P\over c^2}). 
\end{equation}
When the velocity of particles in the gas is non-relativistic, $P\ll \rho c^2$, and we get the familiar Poisson equation $\nabla^2\phi=4\pi G \rho$.

\section{References}

Abbott, B P. 2017, Ann. Phys., 529, Issue 1-2, 1600209

Apostolatos, T., Kennefick, D., Ori, A., \& Poisson, E. 1993, Phys. Rev. D, 47, 5376 

Bardeen, J. M., Press, W. H., \& Teukolsky, S. A. 1972, \apj, 178, 347

Buonanno, A., \& Damour, T. 2000, Phys. Rev. D, 62, 6

Davis, M., Ruffini, R., Press, W. H., \& Price, R. H. 1971, Phys. Rev. Lett. 27, 1466

Davis, M., Ruffini, R., \& Tiomno, J. 1972, Phys. Rev. D, 5, 2932

Detweiler, S. L. 1978, \apj, 225, 687

Detweiler, S. L., \& Szedenits, E. 1979 \apj, 231, 211

Hughes, S. A. 2000, Phys. Rev. D, 61, 084004
 
Hughes, S. A. 2001, Phys. Rev. D, 64, 064004 

Kidder, L. E., Will, C. M., \& Wiseman, A. G. 1993, Phys. Rev.
D 47, R4183 

Lincoln , C. W., \& Will, C. M. 1990, Phys. Rev. D, 42, 4

Misner, C. W., Thorne, K. S., \& Wheeler, J. A., Gravitation (Freeman, San Fransisco 1973)

Nagar, A., Damour, T., \& Tartaglia, A. 2007, Classical and Quantum Gravity, 24, 12

Novikov, I. D.  \& Thorne, K. S. , in Black Holes, eds. C. DeWitt and B. S. DeWitt (Gordon and Breach, New York, 1973)

Oohara, K. I. \& Nakamura, T. 1983, Prog. Theor. Phys. 70, 757; 1983, Phys. Lett. A, 94, 349

Ori, A., \& Thorne, K. S. 2000, Phys. Rev. D, 62, 124022

O'Shaughnessy, R. 2003, Phys. Rev. D, 67, 044004

Pugliese, D., Quevedo, H., \& Ruffini, R. 2011, Phys. Rev. D 83, 2

Peters, P. C., \& Mathews, J. 1963, Phys. Rev. 131, 435

Peters, P. C.1964, Phys. Rev. 136, B1224

Poisson, E. 1993a, Phys. Rev. D, 47, 1497

Poisson, E. 1993b, Phys. Rev. D, 48, 4

Ryan, F. D., 1996, Phys. Rev. D, 53, 3064

Sperhake, U, Berti, E., Cardoso, V., Gonz\'alez, J. A., Br\"ugmann, B., \& Ansorg, M. 2008, 
Phys. Rev. D 78, 064069 

Sundararajan, P. A. 2008, Phys. Rev. D, 77, 12

Teukolsky, S. A. 1973, \apj, 185, 635

Wagoner, R. V., \& Will, C. M. 1976, \apj, 210, 764

Weinberg, S., 1972, Gravitation and Cosmology, Wiley, New York

\end{document}